\documentclass[useAMS,usenatbib]{mn2e}
\usepackage{graphicx,longtable,supertabular}

\begin{document}

\title[Starless cores in IRDCs]
{Isolated starless cores in IRDCs in the Hi-GAL survey}
\author[Wilcock, L.~A. et al.]{L.~A. Wilcock$^{1}$, 
D. Ward-Thompson$^{1}$, J.~M. Kirk$^{1}$, D. Stamatellos$^{1}$, A.
Whitworth$^{1}$, \newauthor C. Battersby$^{2}$, D. Elia$^{3}$,
G.~A. Fuller$^{4}$, A. DiGiorgio$^{3}$, M. J. Griffin$^{1}$, \newauthor S. Molinari$^{3}$, 
P. Martin$^{5}$, J.~C. Mottram$^{6,7}$, N. Peretto$^{8}$, M. Pestalozzi$^{3}$, E. Schisano$^{3}$, \newauthor
H.~A. Smith$^{9}$ \& M.~A. Thompson$^{10}$ \\ 
$^{1}$School of Physics and Astronomy, Cardiff University, Queen's Buildings, 
Cardiff, CF24 3AA, UK \\
$^{2}$ Center for Astrophysics \& Space Astronomy, University of Colorado, 
Boulder, Colorado, 80309, USA \\
$^{3}$ Instituto di Fisica dello Spazio Interplanetario, CNR, via 
Fosso del Cavaliere, I-00133 Roma, Italy \\
$^{4}$ Jodrell Bank Centre for Astrophysics, School of Physics and 
Astronomy, University of Manchester, Manchester, M13 9PL, UK \\
$^{5}$ Canadian Institute for Theoretical Astrophysics, University 
of Toronto, Toronto, Canada, M5S 3H8 \\
$^{6}$ School of Physics, University of Exeter, Stocker Road, Exeter, 
EX4 4QL, UK \\
$^{7}$Leiden Observatory, Leiden University, PO Box 9513, 2300 RA Leiden, The Netherlands \\
$^{8}$  Laboratoire AIM, CEA/DSM-CNRS-Universit\'e Paris Diderot, 
IFRU/Service d'Astrophysique, C.E. Saclay, 
Orme des merisiers, \\ 91191 Gif-sur-Yvette, France \\
$^{9}$ Harvard-Smithsonian Center for Astrophysics, 60 Garden Street, 
Cambridge, MA, 02138, USA \\
$^{10}$ Centre for Astrophysics Research, Science and Technology 
Research Institute, University of Hertfordshire, AL10 9AB, UK \\}
\maketitle

\label{firstpage}

\begin{abstract}
%In a previous paper we identified 1205 infrared dark clouds (IRDCs) that 
%were both Spitzer-dark in the mid-infrared (MIR) and Herschel-bright in 
%the far-infrared (FIR). We identified 972 IRDCs that contained 
%cores, as seen in the FIR at 250~$\mu$m. We placed the cores in an
%evolutionary sequence based on the presence or absence of an embedded 
%point source at different MIR wavelengths. We regarded those without
%embedded sources as the least evolved, and labelled them starless. 

In a previous paper we identified cores within infrared dark clouds (IRDCs). 
We regarded those without embedded sources as the least evolved, and labelled 
them starless. Here 
we identify the most isolated starless cores and model them using a 
three-dimensional, multi-wavelength, Monte Carlo, radiative transfer code. 
We derive the cores' physical parameters and discuss the relation
between the mass, temperature, density, size and the surrounding 
interstellar radiation field (ISRF) for the cores.
The masses of the cores were found not to correlate 
with their radial size or central density.
The temperature at the surface of a core was seen to depend
almost entirely on the level of the ISRF surrounding the core. 
No correlation was found between the temperature at the centre of a core 
and its local ISRF. This was seen to depend, instead, 
on the density and mass of the core.
\end{abstract}


\section{Introduction}
IRDCs were initially discovered by the \textit{MSX} \citep{carey98, egan98} 
and \textit{ISO} \citep{perault96}
surveys as dark regions against the mid-infrared (MIR) background. The 
densest IRDCs may eventually form massive stars \citep{kauffmann10},
and are therefore presumed to represent the earliest observable stage 
of high mass star formation.  

\begin{table*}
\begin{center}
\caption{The physical properties of the modelled cores. 
Column 1 gives the name of the parent cloud as it appears in PF09.
Columns 2 and 3 show the Right Ascension and Declination of the cores in degrees. 
Column 4 is the calculated kinematic distance from the sun to each core (see Section \ref{distance}).
These distances have an uncertainty of $\sim$20\,per\,cent based on how accurately the CO velocity can be found.
Column 5 shows the interstellar radiation field needed to match the model to core's SED, 
given in terms of multiples of the \protect\citet{black94} radiation field. 
Column 6 is the central density of the core. 
Column 7 is the core's semi-major axis, as measured at 250\,$\mu$m. 
Column 8 is the flattening radius of the core and is set to be one-tenth of the semi-major axis. 
Column 9 gives the asymmetry factor, a measure of the eccentricity of the core. 
Column 10 is the mass of each individual core. 
Columns 11 and 12 show the temperatures at the centre and the surface of each core, respectively, 
assuming a viewing angle of $\theta$=0\,\degr. The temperatures and masses of the 
cores are calculated by the model (see text for more details).
The $\chi^2$ values show how the goodness of fit of the model. See text for details.}
\label{coreprop}
\begin{tabular}{ccccccccccccc} \hline
Name of the & RA & Dec & Distance & ISRF & \textit{n}$_{0}$(H$_2$) & $R_{max}$ & $R_0$ & Asymmetry & Mass & 
\multicolumn{2}{l}{Temperature (K)} & $\chi ^2$\\
Parent IRDC & 2000 ($^{\circ}$) & 2000 ($^{\circ}$) & (pc) & & (cm$^{-3}$) & (pc) & (pc) & factor & (M$_{\odot}$) & Centre & 
Surface & \\\hline
305.798$-$0.097 & 199.15 & $-$62.83 & 2900 & 3.8 & 3.0$\times10^4$ & 0.9 & 0.09 & 3.0 & 208 & 11.0 & 22.1 & 1.51\\ 
307.495+0.660 & 202.57 & $-$61.87 & 3600 & 0.7 & 5.0$\times10^3$ & 0.9 & 0.09 & 3.0 & 34.7 & 11.3 & 17.1 & 1.41\\ 
309.079$-$0.208 & 206.24 & $-$62.44 & 3500 & 1.2 & 1.3$\times10^4$ & 0.9 & 0.09 & 3.0 & 90.2 & 10.3 & 18.3 & 2.77\\ 
309.111$-$0.298 & 206.35 & $-$62.52 & 3900 & 5.3 & 1.7$\times10^4$ & 0.7 & 0.07 & 3.0 & 53.8 & 13.5 & 23.9 & 1.94\\ 
310.297+0.705 & 208.33 & $-$61.28 & 4400 & 2.3 & 5.6$\times10^4$ & 0.9 & 0.09 & 1.3 & 75.3 & 10.1 & 20.6 & 2.48 \\ 
314.701+0.183 & 217.28 & $-$60.44 & 3500 & 1.6 & 9.5$\times10^4$ & 0.6 & 0.06 & 2.0 & 103 & 8.80 & 19.0 & 1.83\\ 
318.573+0.642 & 223.84 & $-$58.41 & 2900 & 1.8 & 3.4$\times10^4$ & 0.5 & 0.05 & 3.0 & 40.5 & 10.5 & 19.6 & 0.75\\ 
318.802+0.416 & 224.43 & $-$58.51 & 2900 & 4.6 & 5.5$\times10^3$ & 0.6 & 0.06 & 3.0 & 11.3 & 16.5 & 23.8 & 2.16\\ 
318.916$-$0.284 & 225.26 & $-$59.07 & 2500 & 2.2 & 5.7$\times10^4$ & 0.4 & 0.04 & 3.0 & 34.7 & 10.2 & 20.0 & 0.72\\ 
321.678+0.965 & 228.59 & $-$56.61 & 2100 & 0.6 & 2.5$\times10^4$ & 0.6 & 0.06 & 3.0 & 51.4 & 8.80 & 16.1 & 4.04\\ 
321.753+0.669 & 228.99 & $-$56.83 & 2000 & 1.6 & 1.5$\times10^3$ & 0.8 & 0.08 & 3.0 & 7.31 & 15.6 & 20.4 & 2.32\\ 
322.334+0.561 & 229.99 & $-$56.61 & 3500 & 2.0 & 1.3$\times10^4$ & 0.9 & 0.09 & 3.0 & 90.2 & 11.5 & 20.2 & 1.45\\ 
322.666$-$0.588 & 231.65 & $-$57.39 & 3500 & 1.3 & 1.3$\times10^5$ & 1.0 & 0.10 & 1.0 & 63.9 & 8.10 & 19.1 & 4.35 \\ 
322.914+0.321 & 231.11 & $-$56.49 & 2100 & 3.7 & 2.0$\times10^5$ & 0.4 & 0.04 & 1.2 & 17.9 & 10.3 & 22.4 & 2.14 \\ 
326.495+0.581 & 235.97 & $-$54.20 & 2700 & 10.4 & 1.2$\times10^5$ & 0.4 & 0.04 & 2.9 & 69.6 & 12.0 & 26.2 & 1.02\\ 
326.620$-$0.143 & 236.91 & $-$54.70 & 3000 & 2.2 & 4.5$\times10^3$ & 1.0 & 0.10 & 3.0 & 38.8 & 14.1 & 20.8 & 1.58\\ 
326.632+0.951 & 235.77 & $-$53.83 & 2600 & 4.0 & 9.5$\times10^4$ & 0.3 & 0.03 & 2.4 & 17.5 & 10.4 & 23.0 & 10.12\\ 
326.811+0.656 & 236.32 & $-$53.95 & 2800 & 8.5 & 2.0$\times10^5$ & 0.3 & 0.03 & 1.4 & 57.4 & 11.3 & 25.5 & 2.07 \\ 
328.432$-$0.522 & 239.72 & $-$53.84 & 3100 & 2.8 & 6.7$\times10^3$ & 0.8 & 0.08 & 2.8 & 29.6 & 14.2 & 21.6 & 2.18\\ 
329.403$-$0.736 & 241.19 & $-$53.37 & 4500 & 3.0 & 1.4$\times10^4$ & 1.0 & 0.10 & 2.0 & 70.1 & 12.7 & 21.7 & 2.51 \\ \hline
\end{tabular}
\end{center}
\end{table*}

Some IRDCs contain cold, compact condensations called infrared dark 
cores; also referred to as `clumps' (e.g.
\citealt{battersby10}) or `fragments' (e.g. \citealt{peretto09}). 
The most massive of these objects are believed 
to be the high mass equivalent of low-mass prestellar cores 
(\citealt{wardthompson94,carey00,redman03,garay04}). 

Observations of IRDCs have shown them to have low temperatures 
(T$<$25\,K; e.g. \citealt{egan98, teyssier02}), high densities 
(n$_H$$>$$10^5 $\,cm$^{-3}$; \citealt{egan98,carey98,carey00}) 
and masses between $\sim$10$^2$-10$^4$\,M$_{\odot}$
\citep{rathborne05a,ragan09}.

Within the IRDCs, infrared dark cores have been observed with radii 
ranging from 0.04 to 1.6\,pc, masses from 10 to 1000\,M$_{\odot}$ 
and densities from 10$^3$ to 10$^7$\,cm$^{-3}$ 
\citep{garay04,ormel05,rathborne05a,rathborne06,swift09,zhang11}. 
This is consistent with them being the 
precursors to hot molecular cores, the next observable stage 
in high mass star formation \citep{rathborne06}.

Using CO and CS velocities, kinematic distances have been 
calculated for several hundred IRDCs and, by extension, their cores
\citep{simon061,jackson08}. The distribution of IRDCs in the first 
Galactic quadrant differs from those in the fourth quadrant. In
the first quadrant IRDCs have a mean galacto-centric distance of $\sim$5\,kpc \citep{jackson08}, whereas 
those in the fourth quadrant are typically found at $\sim$6\,kpc \citep{peretto10b}. This 
suggests that many IRDCs may be located in the Scutum-Centaurus 
spiral arm. The association of IRDCs to a spiral arm lends weight 
to the theory that IRDCs are the birthplace of high mass stars 
\citep{jackson08}.

Cores within IRDCs can be separated into different classes based 
on observational parameters. Starless cores, also called 
`quiescent cores', are the youngest subset of infrared dark cores. 
They are not yet
undergoing any form of active star formation and so contain no MIR 
activity \citep{chambers09}. As the cores evolve they leave the 
starless phase and
begin to exhibit tracers of star formation. First 24-$\mu$m emission, 
and then PAH 8-$\mu$m emission
(see \citealt{catpaper} for more details). When the cores 
have reached the latter stage they are thought to contain a 
hyper- or ultra-compact HII region \citep{chambers09,battersby10}.

In a previous paper \citep{catpaper}, we used data from the \textit{Herschel}
Infrared Galactic Plane Survey (Hi-GAL; \citealt{higalb,higala}) to observe 
the IRDCs from \citet{peretto09} -- hereafter PF09 -- at FIR 
wavelengths. We identified 1205 IRDCs that were simultaneously 
Spitzer-dark in the MIR and Herschel-bright in the FIR, and found that
972 of them contained identifiable cores, as defined by peaks in their
FIR emission at 250~$\mu$m. We classified the cores depending on the
presence or absence of an embedded MIR source at 8- or 24-$\mu$m.
We identified the youngest, least-evolved cores as those which contained
no embedded MIR point sources, and labelled them starless. We found 170
IRDCs containing starless cores. 

Previous study has shown that high mass star formation is unlikely to have already occurred
in starless cores \citep{wilcock11}. However, it should be noted that, if the column density were high enough, 
it is possible for an IRDC to absorb the MIR emission from more evolved cores and appear starless 
despite the presence of embedded protostars \citep{pavlyuchenkov11}.

In this paper we identify the most isolated of these 170 starless cores,
based on their appearance at FIR wavelengths. We find 20 starless cores 
that satisfy our conditions for being the most isolated. We then model these
20 cores to determine their physical parameters.

\begin{figure*}
\includegraphics[angle=0,width=175mm]{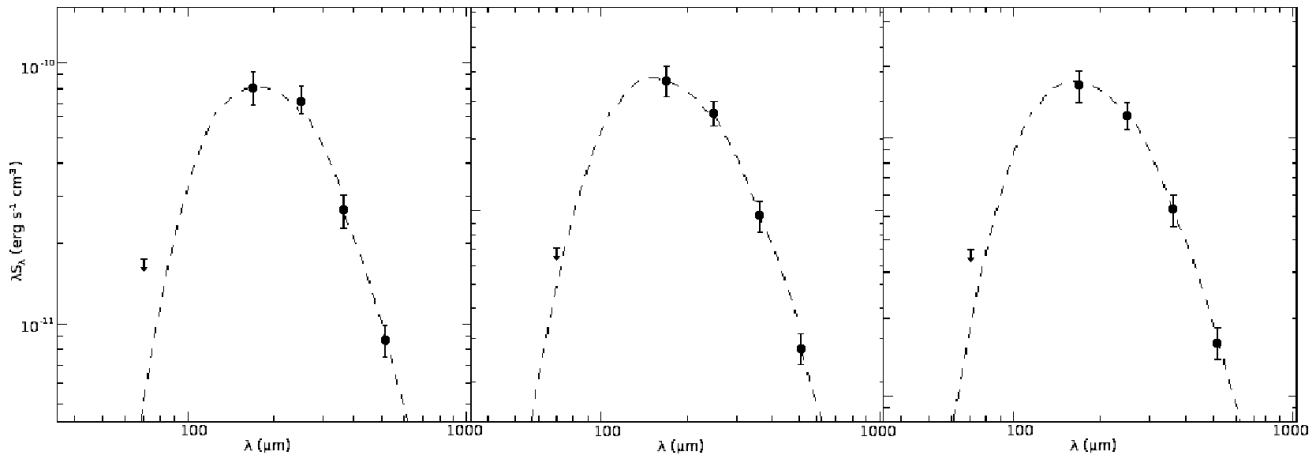}
\caption{Example SEDs for cores 307.495+0.660, 318.573+0.642 and 
322.334+0.561. The points show the observed fluxes at each of our
wavelengths and the model SED is shown as a dashed 
line. The 70-$\mu$m data provide upper-limits only.} 
\label{SED}
\end{figure*}

\section{Data}\label{data}

The \textit{Herschel}\footnote{\textit{Herschel} 
is an ESA space observatory with science instruments provided by 
European-led Principal Investigator consortia and 
with participation from NASA.} Space Observatory was launched in May 2009. 
\textit{Herschel} carries three science instruments.
These are: the Spectral and Photometric Imaging Receiver 
(SPIRE, \citealt{spire}); the 
Photodetector Array Camera and Spectrometer (PACS, \citealt{pacs}); 
and the Heterodyne Instrument for the Far Infrared (HIFI, 
\citealt{hifi}). \textit{Herschel}
is capable of observing in the FIR between 55 and 671~$\mu$m. 
We only use data here from SPIRE and PACS.

The data used in this paper were taken as part of the \textit{Herschel} 
Infrared Galactic Plane Survey (Hi-GAL), an Open Time Key
Project of the \textit{Herschel} Space Observatory \citep{higalb,higala}. 
Hi-GAL aims to perform a survey of the Galactic Plane 
using the PACS and SPIRE instruments. The two are used in parallel mode to 
map the Milky Way Galaxy simultaneously at five 
wavelengths (70, 160, 250, 350 and 500\,$\mu$m), with resolution up to 
5\arcsec{} at 70\,$\mu$m.

PACS data reduction was performed using the \textit{Herschel} Interactive 
Pipeline Environment (HIPE; \citealt{hipe}), with some
additions described by \citet{pacs}. The standard deglitching and cross-talk 
correction were not used and custom procedures were 
written for drift removal \citep{traficante11}. 

SPIRE data processing used the standard processing methods \citep{spire}, 
with both standard deglitching and drift removal. 
In both cases, the ROMAGAL Generalised Least Squares algorithm 
\citep{traficante11} was used to produce the final maps. A more 
detailed discussion of the data reduction process is given by 
\citet{traficante11}.

We also make use of 8-$\mu$m and 24-$\mu$m data from 
\textit{Spitzer}\footnote{The \textit{Spitzer}
Space  Telescope was operated by the Jet Propulsion Laboratory
at the California Institute of Technology under a contract with NASA.}
that were taken as part of the Galactic Legacy Infrared Mid-Plane 
Survey Extraordinaire (GLIMPSE; \citealt{glimpse}) and MIPSGAL \citep{mipsgal}, respectively. 
We used the mosaics available from the \textit{Spitzer} Science Centre
to create 8- and 24-$\mu$m maps of the region 
$300\,\degr$$\le$$l$$\le$$330\,\degr$ and $|b|$$\le$$1\,\degr$.

\begin{figure*}
\includegraphics[angle=-90,width=85mm]{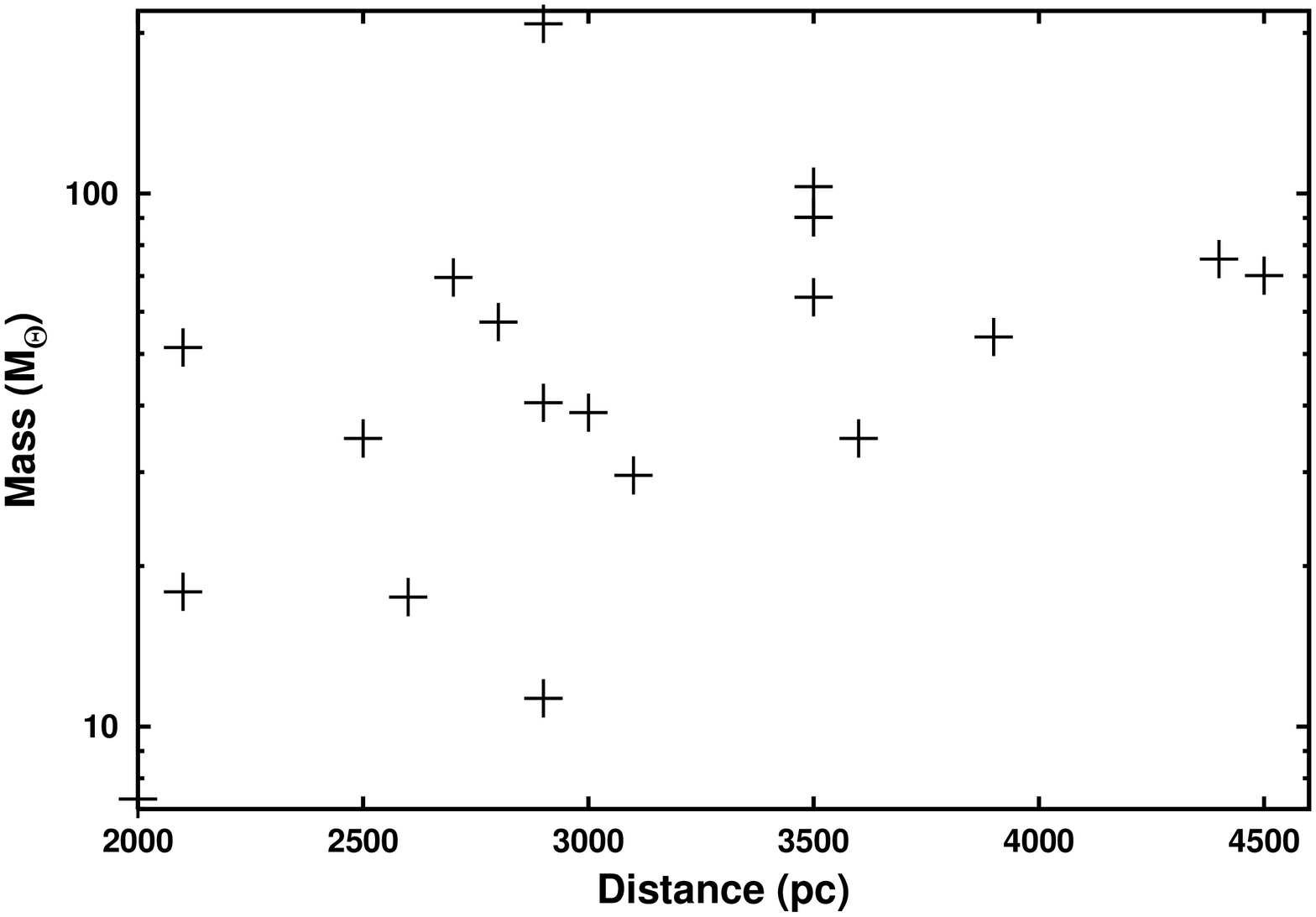}
\includegraphics[angle=-90,width=85mm]{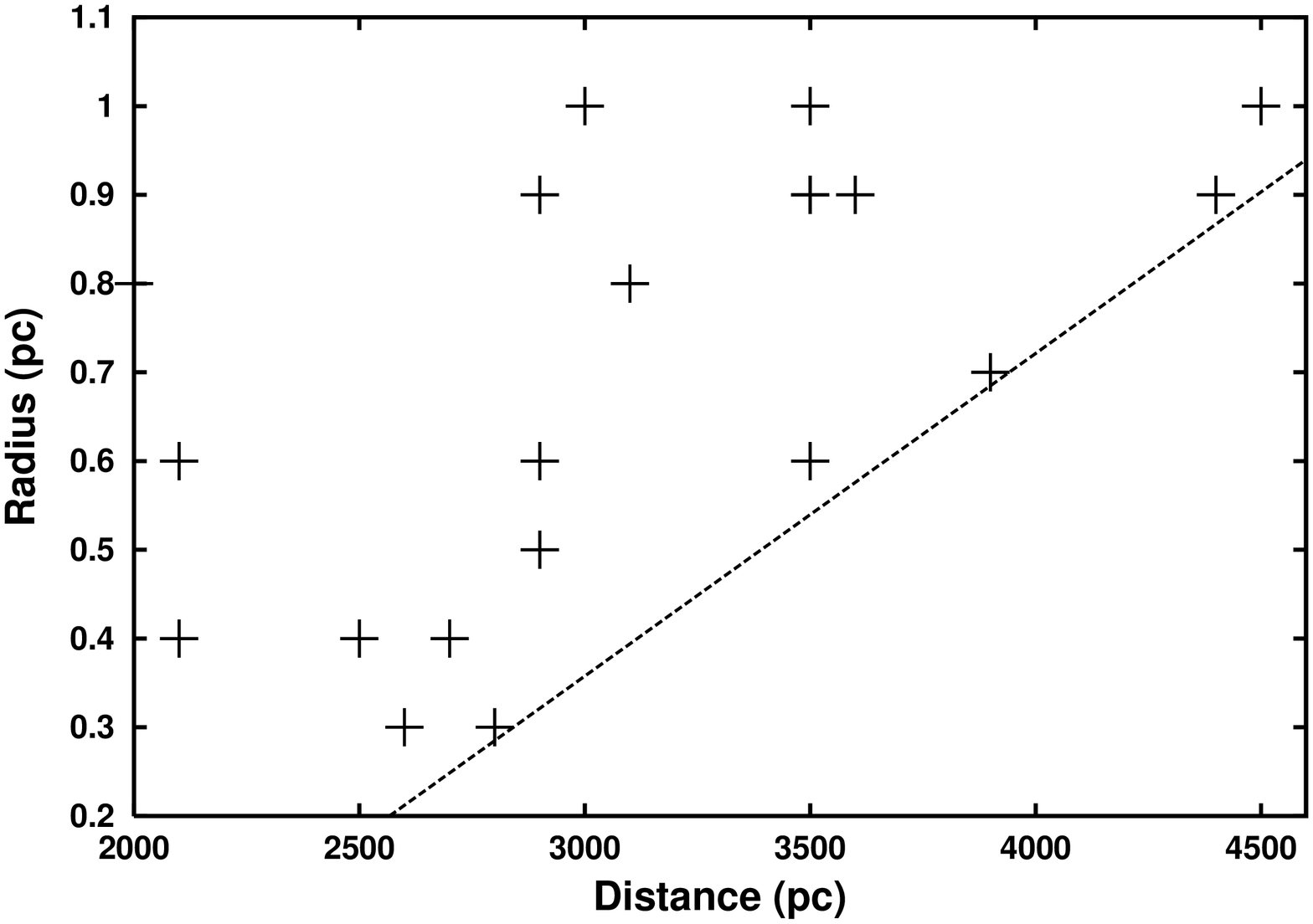}
\caption{Mass (left) and radius (right) versus core helio-centric
distance. The dashed line shows 
the cutoff point for a core's radius -- a core with 
a radius below this line will not have been selected for
modelling due to our `isolated' criterion (see text).} 
\label{dist_rad_mass}
\end{figure*}

\section{Source Selection and Modelling}

\subsection{Source sample} \label{findingirdcs}

We took as our initial sample the 170 IRDCs with starless cores identified
by \citet{catpaper}. A Gaussian profile was fitted to the 250-$\mu$m intensity 
of each core. This was used to determine the full-width half-maximum (FWHM) of 
the core. Each starless core was also studied at 160 and 500~$\mu$m to 
determine if it was isolated. %Isolated cores are best suited to our model.
Our model is best suited to isolated cores because isolated cores are more likely 
to have the simple internal structures.

For a core to be deemed isolated, it has to be unconfused by a neighbouring
core within the IRDC. We defined this by saying that there
must be a minimum depth of magnitude greater than 3~$\sigma$ between the 
peak of one core and that of any neighbouring core. This must be true at all FIR 
wavelengths (160, 250, 350 and 500~$\mu$m). Of the 170 IRDCs with starless cores, 
only 20 cores were found to be truly isolated according to our rigorous definition. 
These cores are listed in Table~\ref{coreprop}. 

\subsection{Distances}\label{distance} 

The distances to our infrared dark cores had not been previously calculated. 
The galacto-centric distance to each core was therefore estimated using 
the \citet{brand93} velocity curve:
\begin{equation}
\frac{\theta}{\theta_0} = a_1 \left(\frac{d_g}{d_{g,0}}\right)^{a_2}+a_3
\end{equation}
where $a_1=1.0069$, $a_2=0.0363$ and $a_3=0.0065$ in the fourth quadrant 
of the Galactic Plane. 
$\theta$ is the core's orbital velocity and $d_g$ is its galacto-centric 
distance.
$\theta_0$ and $d_{g,0}$ are the solar equivalents. Canonical values 
of 8.5\,kpc and 220\,kms$^{-1}$
were used for $d_{g,0}$ and $\theta_0$ respectively. CO velocities 
were taken from \citet{dame87}. This method is only 
accurate up to 8\,kpc \citep{brand93}.

From this, the heliocentric distance was calculated using
\begin{equation}
d_g = \left( d_h^2\,cos^2b + d_{g,0}^2 - 2\,d_{g,0}\,d_{h}\,cos(b)\,cos(l) 
\right) ^{\frac{1}{2}}
\end{equation}
where $d_h$ is the core's heliocentric distance and $b$ and $l$ are the 
Galactic latitude and longitude of the core.
This is solved as a quadratic equation and results in two possible 
distances for each core --- one on the near side of the Galaxy and
one on the far side. 

The near/far distance ambiguity exists in all cases. However, each 
of our cores had to show up in absorption at 8~$\mu$m to be considered a
candidate IRDC in the first place (PF09).
Therefore, most of the MIR-emitting material must be behind our cores, 
rather than in front of them, or 
absorption would not be evident. For this reason, we assume that 
the near distance is correct for all of the cores. The calculated 
distance to each core is listed in Table~\ref{coreprop}. 

\begin{figure*}
\includegraphics[angle=-90,width=80mm]{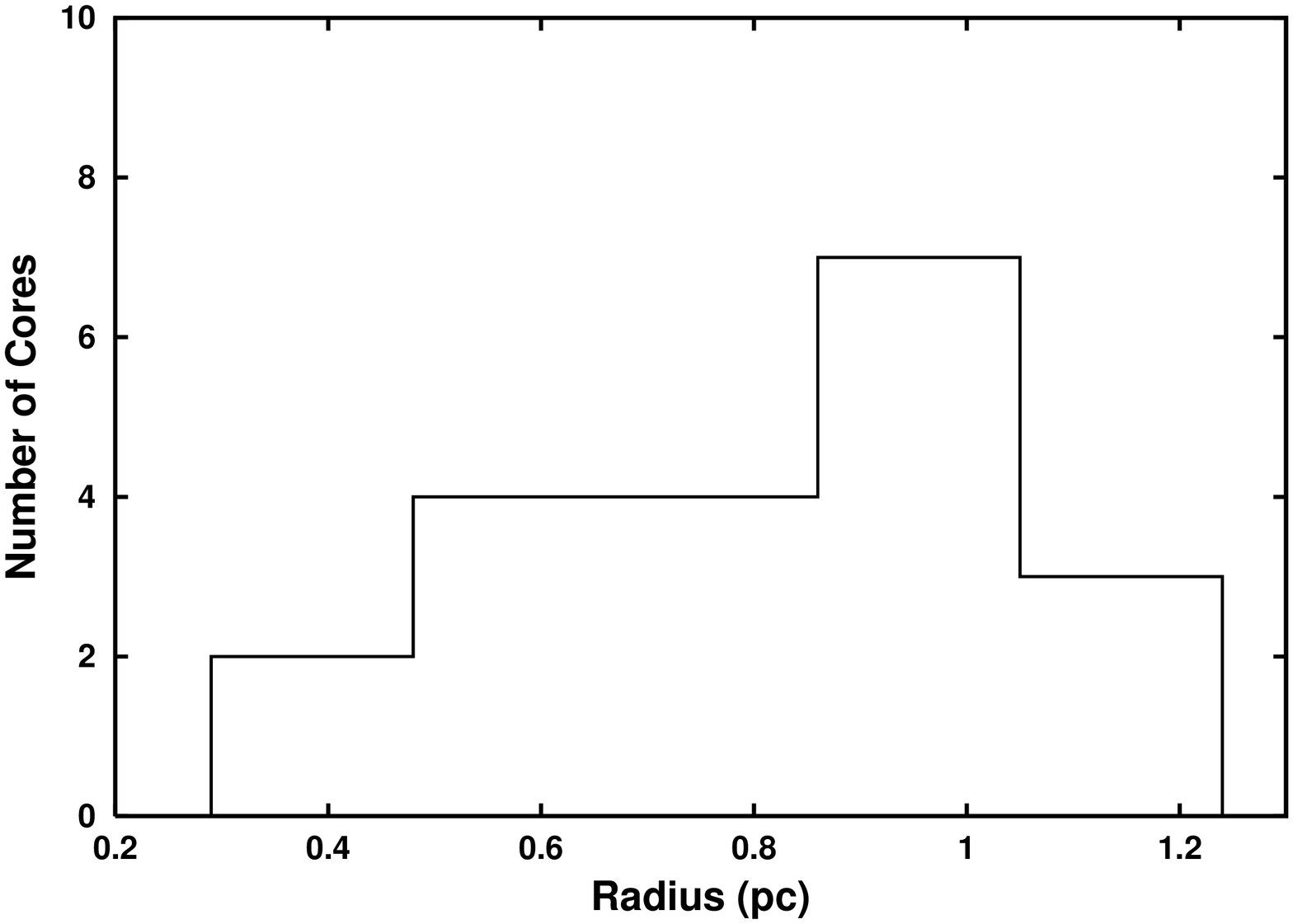}
\includegraphics[angle=-90,width=80mm]{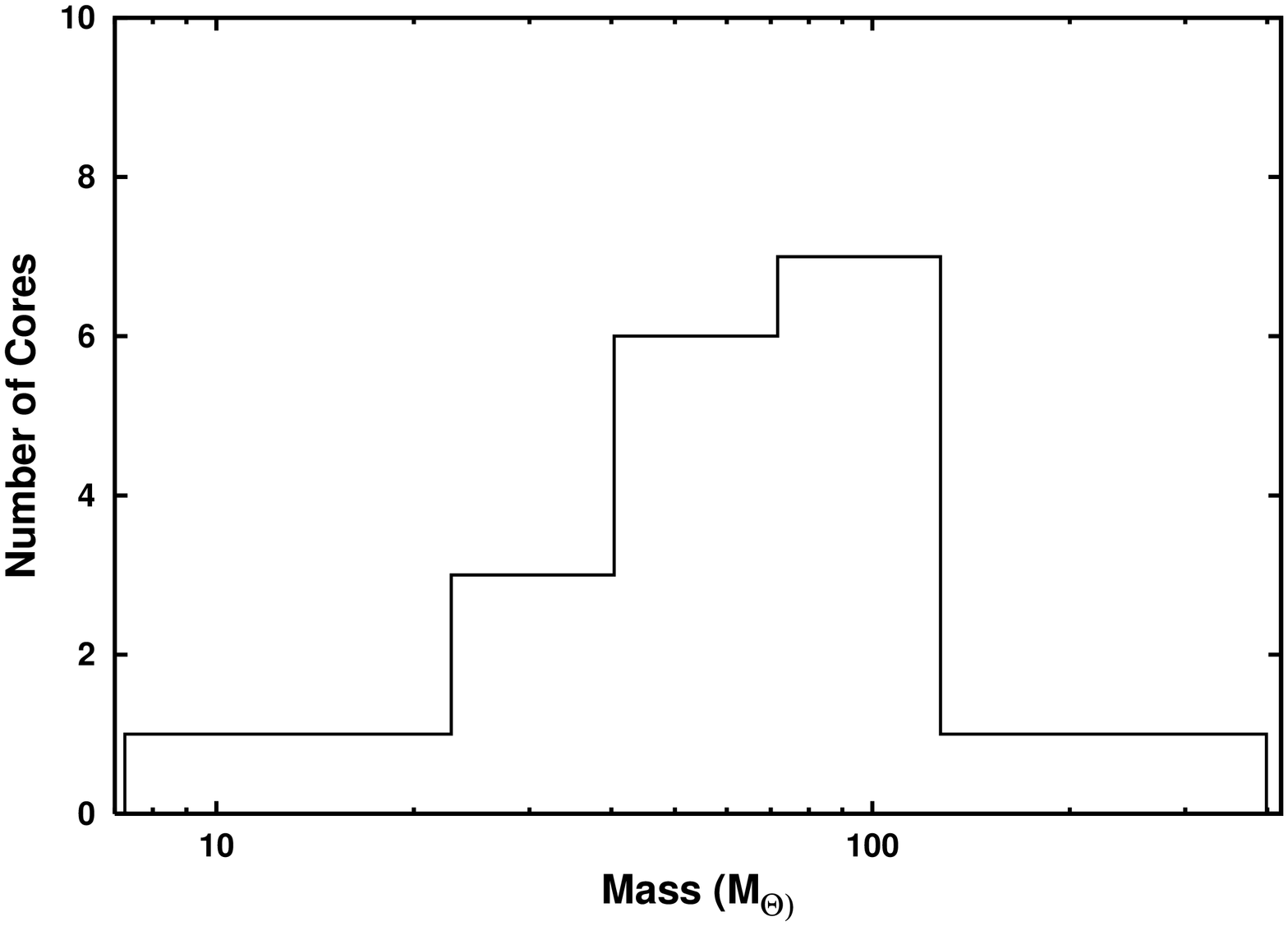}
\includegraphics[angle=-90,width=80mm]{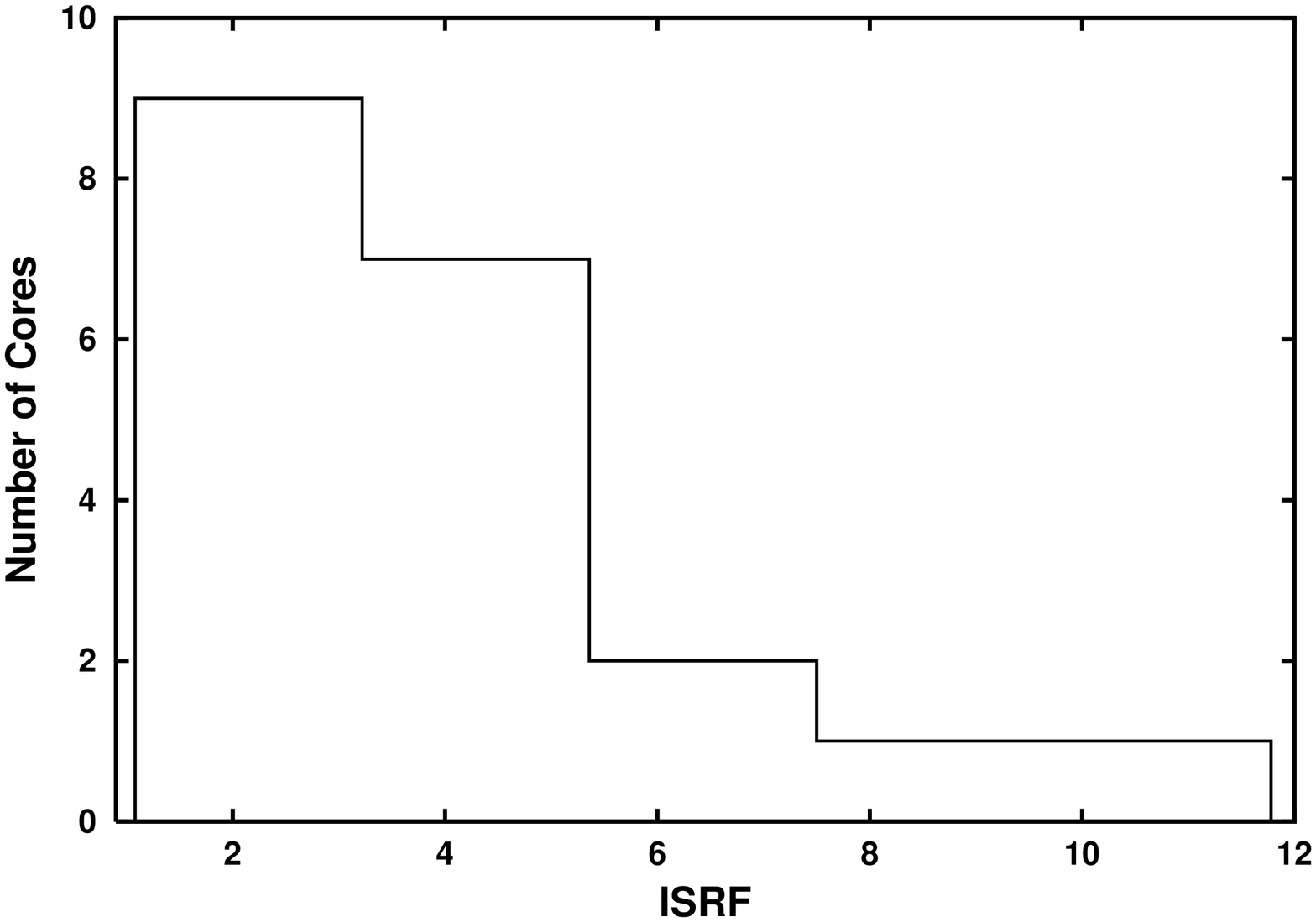}
\includegraphics[angle=-90,width=80mm]{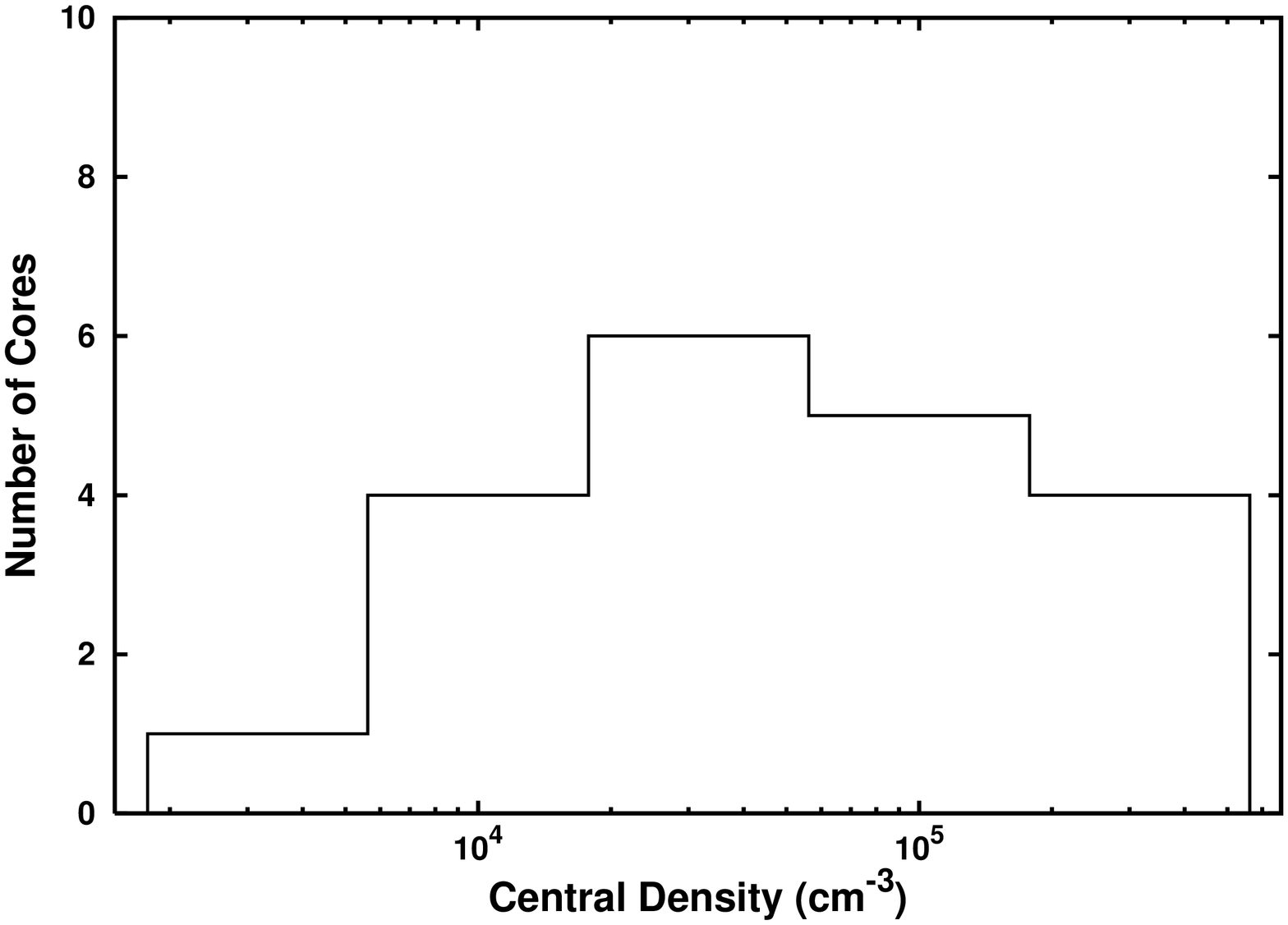}
\includegraphics[angle=-90,width=80mm]{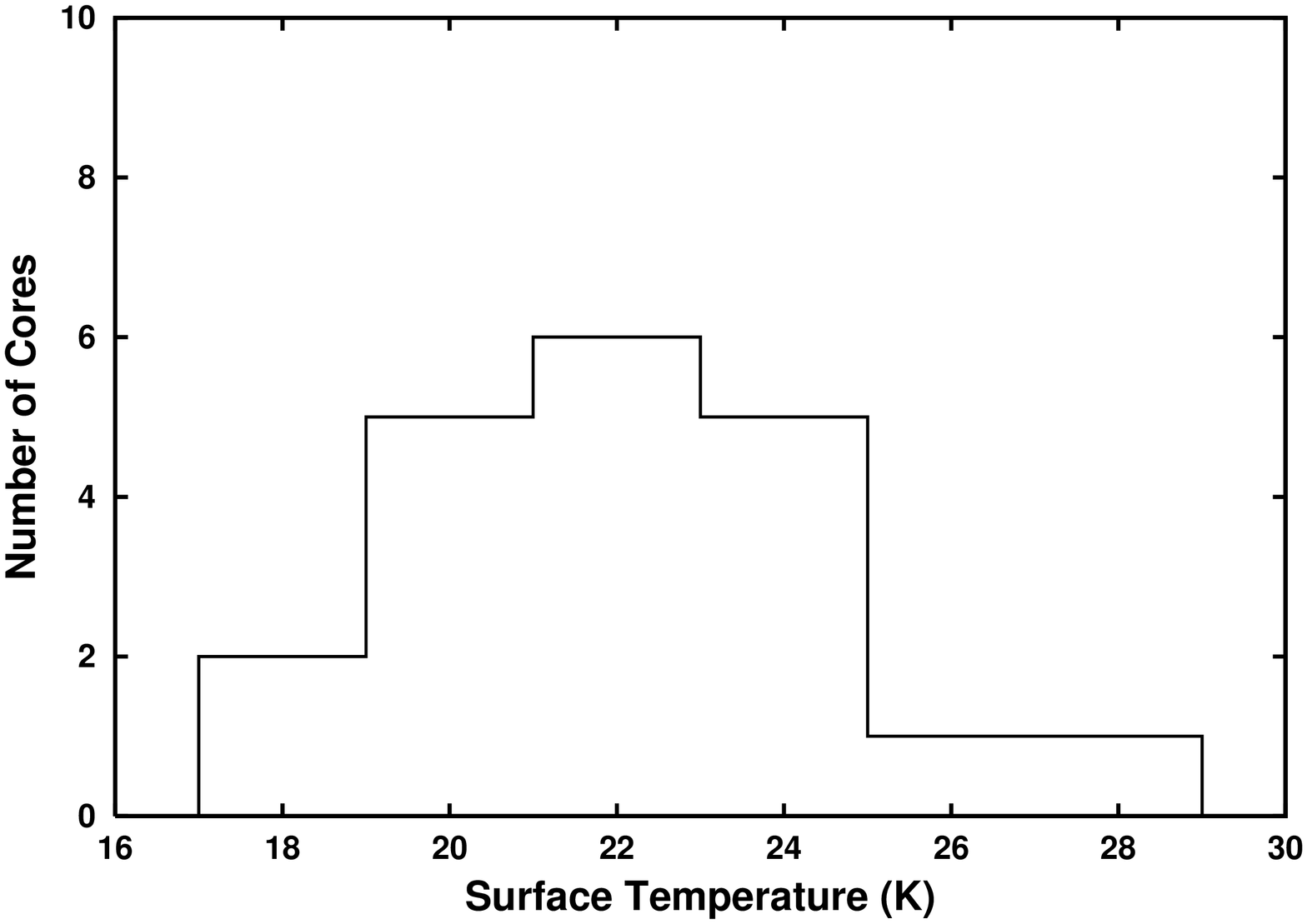}
\includegraphics[angle=-90,width=80mm]{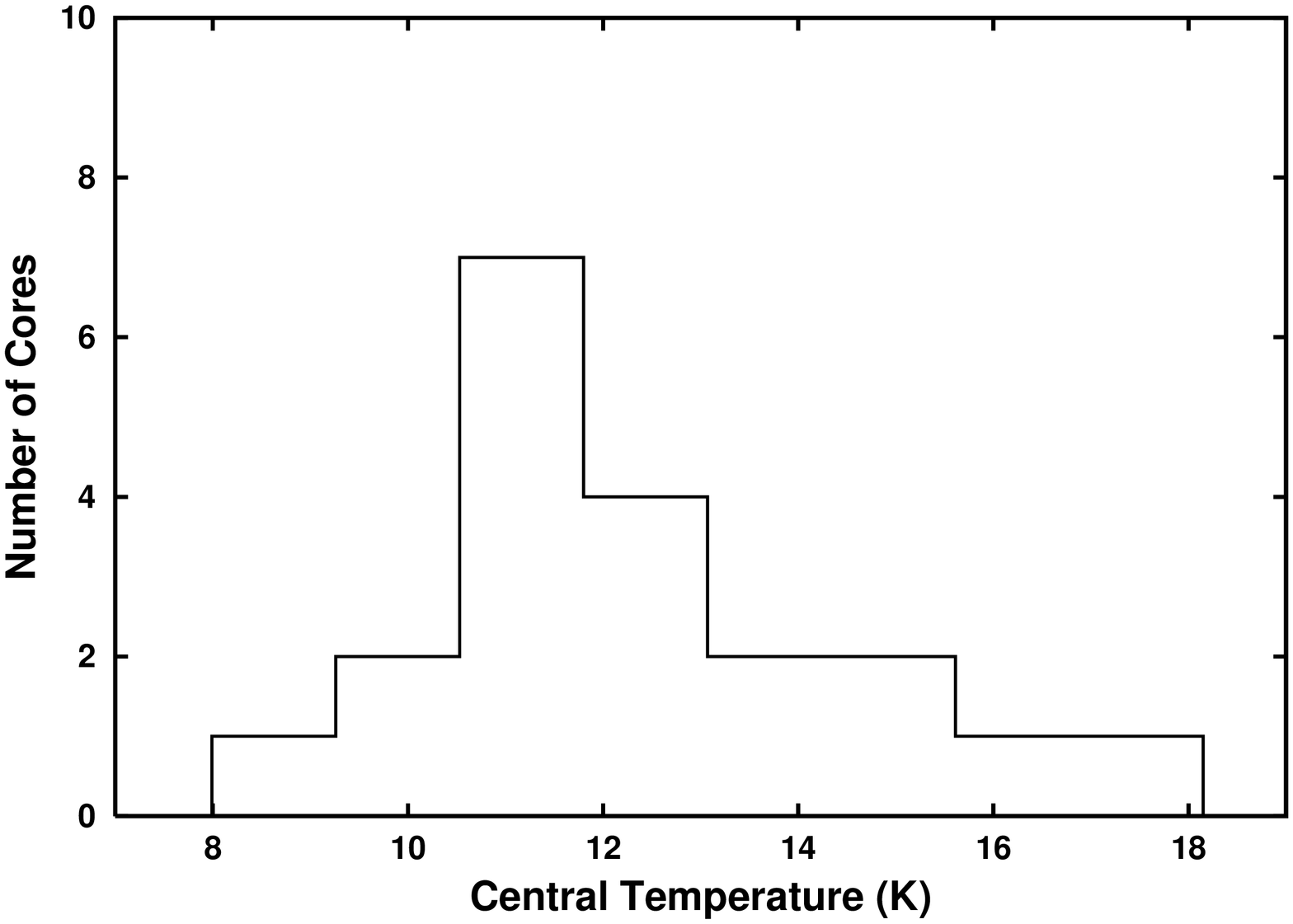}

\caption{Histograms of (top row, left to right) core radius, core mass, 
(middle row) the ISRF surrounding the core and the core's 
central density and (lower row) the surface and central 
temperatures found in the cores. Note that the x-axes for mass and central
density use a logarithmic scale.}
\label{hist}
\end{figure*}

Fourth quadrant IRDCs have a typical galacto-centric distance of 6\,kpc 
\citep{jackson08} and a typical heliocentric distance of 4\,kpc
\citep{peretto10b}. The mean heliocentric distance to our isolated cores 
is 3.1\,kpc, closer than 4\,kpc. This is due to a selection bias 
in our isolation criterion. It can be seen in 
Figure~\ref{dist_rad_mass}, where we compare the masses 
(see discussion below) and 
radii of the cores to their distance. Only cores with a high mass or 
large radius were seen as isolated at greater distances. 
This is due to the annular resolution of \textit{Herschel}.
Using our lowest 
data-points, a cutoff is extrapolated This is shown as a dashed line in 
Figure~\ref{dist_rad_mass}. Below this cutoff a core is unlikely to be 
selected for modelling due to our isolation criterion.

\subsection{Radiative Transfer Modelling}\label{modelling}

Our 20 isolated, starless cores were modelled using \textsc{Phaethon} 
\citep{stamatellos03, stamatellos05, stamatellos10}.
\textsc{Phaethon} is a 3D 
Monte Carlo radiative transfer code. The code uses luminosity packets 
to represent the ambient
radiation field in the system. These packets are injected into the 
system where they interact (are absorbed, re-emitted or
scattered) stochastically. 

The input variables of the code are the density profile, the strength 
of the interstellar radiation field (ISRF), the dust
properties of the system, the size of the core and its geometry (i.e. 
spherical, flattened or cometary). The code calculates the
temperature profile of the system as well as spectral energy 
distributions (SEDs -- see Figure \ref{SED}) and intensity maps, 
at different wavelengths and viewing angles --- see Appendix \ref{egcores}.

\textsc{Phaethon} uses \citet{ossenkopf94} opacities for a standard \citet{mathis77} grain mixture 
of 52\,per\,cent silicate and 47\,per\,cent graphite with grains that have coagulated and accreted thin ice mantles over 
$10^5$\,years at densities of 10$^6$\,cm$^{-3}$. Further details of the dust model used can be found in
\citet{stamatellos03,stamatellos05} and \citet{stamatellos10}.

All the cores showed some measure of eccentricity in the observations 
and so were modelled with a flattened geometry --- see
\citet{stamatellos04, stamatellos10} for details. In this case the 
density profile is given by:

\begin{equation}\label{densityeqn}
n \left( r, \theta \right) = n_0 \left( {\rm H}_2 \right) 
\frac{1 + A \left( \frac{r}{R_0} \right) ^2 \left[ \sin (\theta) 
\right] ^p }{\left[ 1 + \left( \frac{r}{R_0} \right) ^2 \right] ^2} ,
\end{equation}

\noindent
where \textit{r} is the radial distance, $\theta$ is the polar 
angle and $R_0$ is the flattening radius (i.e. the radial distance 
for which the central density is approximately constant) and was set to one tenth of the core's semi-major axis, $R_{max}$. 
\textit{n}$_0$(H$_2$) is the central density, which is controlled as an 
input variable. \textit{A} is the asymmetry factor and controls the equatorial 
to polar optical depth ratio and determines how flattened the 
core is (i.e. its eccentricity). \textit{p} determines how quickly the optical depth changes 
from equator to pole, and was set to 2. These values for $R_0$ and \textit{p} were chosen as they have been 
shown previously to give sensible results when modelling cold cores \citep[e.g.][]{stamatellos10,wilcock11}. By restricting these 
parameters to sensible estimates a single model may be fitted to each core.

The size and shape of each modelled core was set to match the observed core at 250\,$\mu$m. 
%The FWHM of the major axis of each observed core was used as the model core's semi-major axis, $R_{max}$. 
$R_{max}$ was set equal to the FWHM of the major axis of each observed core.
The $R_{max}$ of each core is listed in Table \ref{coreprop}. 
The FWHM of the cores have an error of $\pm 15$\,per\,cent, due primarily
to uncertainties in the background levels. 
Contours from the 250\,$\mu$m observations were overlaid onto images of the modelled core. The asymmetry factor 
of each model was manually varied until the modelled shape visually matched that of the observed core.

To test the goodness of fit of the modelled morphology to the observed morphology we compared the 
eccentricity and the FWHM. For the eccentricity, a percentage difference of less than 20\,per\,cent was considered a 
good fit. The eccentricities of the observed and modelled cores are given in Table \ref{ecc}. The FWHM of the 
modelled core is expected to agree with the FWHM of the observed core within the prescribed errors of $\sim$15\,per\,cent. 
The FWHM of the observed and modelled cores at 250\,$\mu$m are given in Table \ref{fwhm}. Differences between the 
model and observed morphology are mainly due to the assumption that the observed cores were perfect ellipses and the 
difficulties of fitting such a model by-eye.

The flux density, integrated over twice the FWHM of each core using an elliptical aperture, was measured in all 
five FIR maps and an SED was plotted. These flux densities have been background subtracted, where the 
background was defined using an off-cloud, elliptical aperture.

The central density, \textit{n}$_0$(H$_2$), and the ISRF incident on the cores were manually varied until the 
output model's SED matched the observed data. The ISRF is 
taken to be a multiple of a modified version of the 
\citet{black94} radiation field, which gives a good 
approximation to the radiation field in the solar neighbourhood. The final values for 
\textit{n}$_0$(H$_2$) and the incident ISRF have an uncertainty of $\pm15$\,per\,cent. This is based
on a by-eye estimation of how much they can be varied until the model SED no longer fits the 
observed flux densities. Examples of the 
model SEDs are shown in Figure~\ref{SED} as a dotted line. The 
final values for \textit{n}$_0$(H$_2$) and the incident ISRF 
are listed in Table~\ref{coreprop}.

To test the goodness of fit of the modelled SED to the observed SED the $\chi ^2$ value was calculated for each core.
Only the flux densities from 160--500\,$\mu$m were taken into account when calculating $\chi ^2$. The 70\,$\mu$m point was 
only used as an upper limit and was ignored in this calculation. $R_{max}$ and the asymmetry factor were set through 
visual observations. As such, only the ISRF and central density were varied when matching the model SED to the observed SED.
There are four data points in each SED and two free parameters, meaning there are two degrees of freedom in our SED fit. 
We aim for a 5\,per\,cent significance level and perform a two-sided test. For a 5\,per\,cent significance level, with two 
degrees of freedom, $\chi ^2$ must be between 0.051 and 7.378 \citep[e.g.][]{rouncefield93} for the model to be deemed a `good' fit 
to the data. The final $\chi ^2$ values for each model are listed in Table \ref{coreprop}.

For example, we can look at the core in 307.495+0.660, the SED of which is shown in Figure \ref{SED} and images of the
observed and modelled core are shown in Appendix \ref{egcores}. The model created for the core in 307.495+0.660 shows 
absorption at 8 and 70\,$\mu$m, comparable to the absorption seen in the observations. The size 
and shape of the model fits with the contours from the observed data at all wavelengths (see Tables \ref{ecc} and \ref{fwhm}).
The model does, however, fit slightly better at longer 
wavelengths where the observed core appears as a more typical ellipse. The observed FWHM of the core is 55\,\arcsec at
160\,$\mu$m and 67\,\arcsec at 500\,$\mu$m. The model core has a FWHM of 65\,\arcsec at 160\,$\mu$m and 62\,\arcsec at 
500\,$\mu$m, in agreement with the observed values. The eccentricity of the modelled cores shows only a 13\,\% difference 
from the eccentricity of the observed core at 250\,$\mu$m. The modelled SED fits well with the observed SED at
all wavelengths and has a $\chi^2$ of 1.41, well within our boundaries for a 5\% significance level. The model is in agreement
with the observed flux densities at 160, 250, 350 and 500\,$\mu$m and below the upper limit set at 70\,$\mu$m.

Nineteen of the twenty modelled cores met the criteria for a good fit to the morphology and SED of the observed cores.
The core in 326.632+0.951, however, did not meet our criteria for a good fit. This core has a $\chi ^2$ of 10.21 and the 
morphology of the model showed a 50\,per\,cent difference from the observed core at 250\,$\mu$m.
It was, nonetheless, the best fit possible with \textsc{Phaethon}. This may be indicating that this core has a more complex
morphology than the model allows.

\begin{figure*}
\includegraphics[angle=-90,width=85mm]{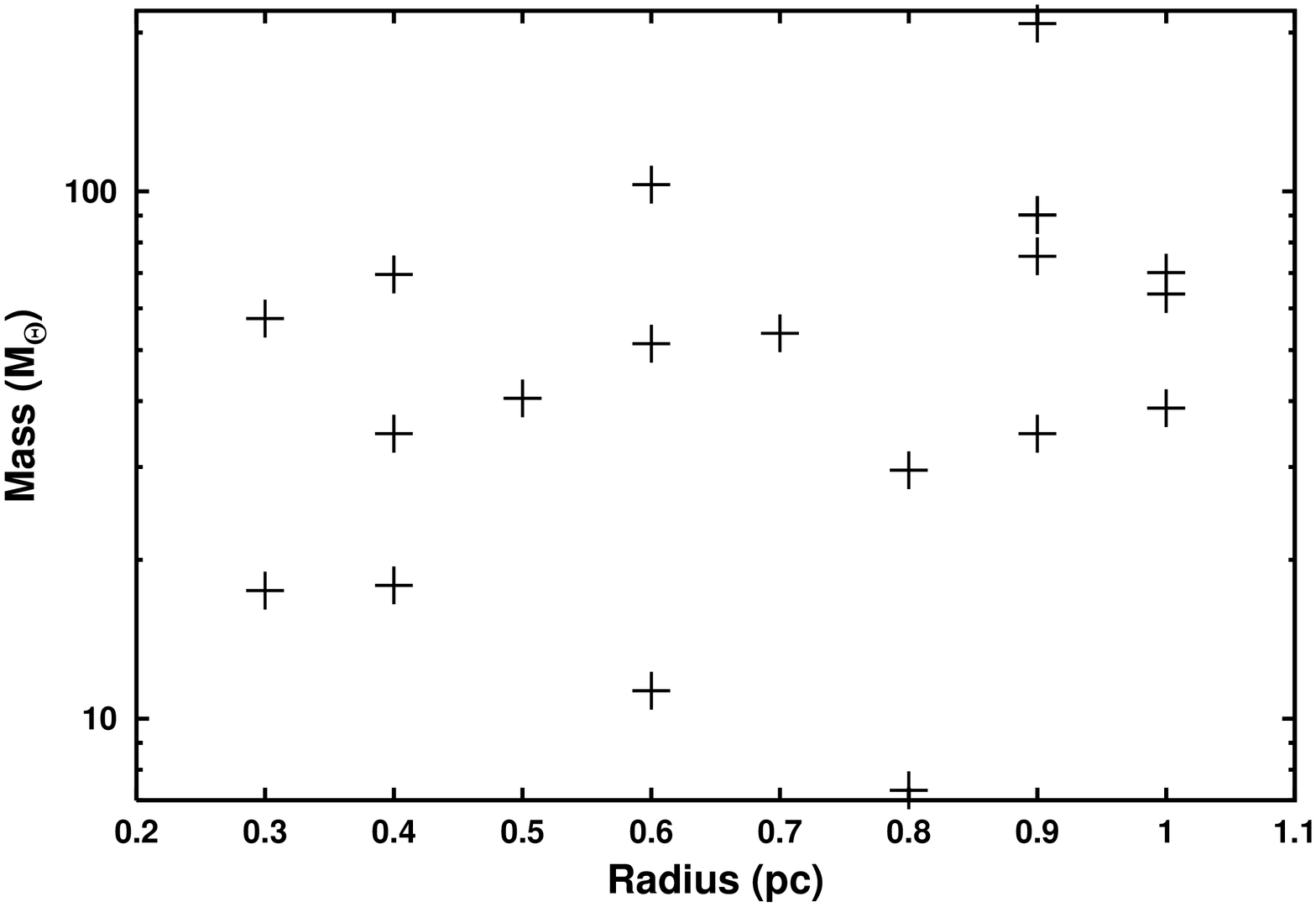}
\includegraphics[angle=-90,width=85mm]{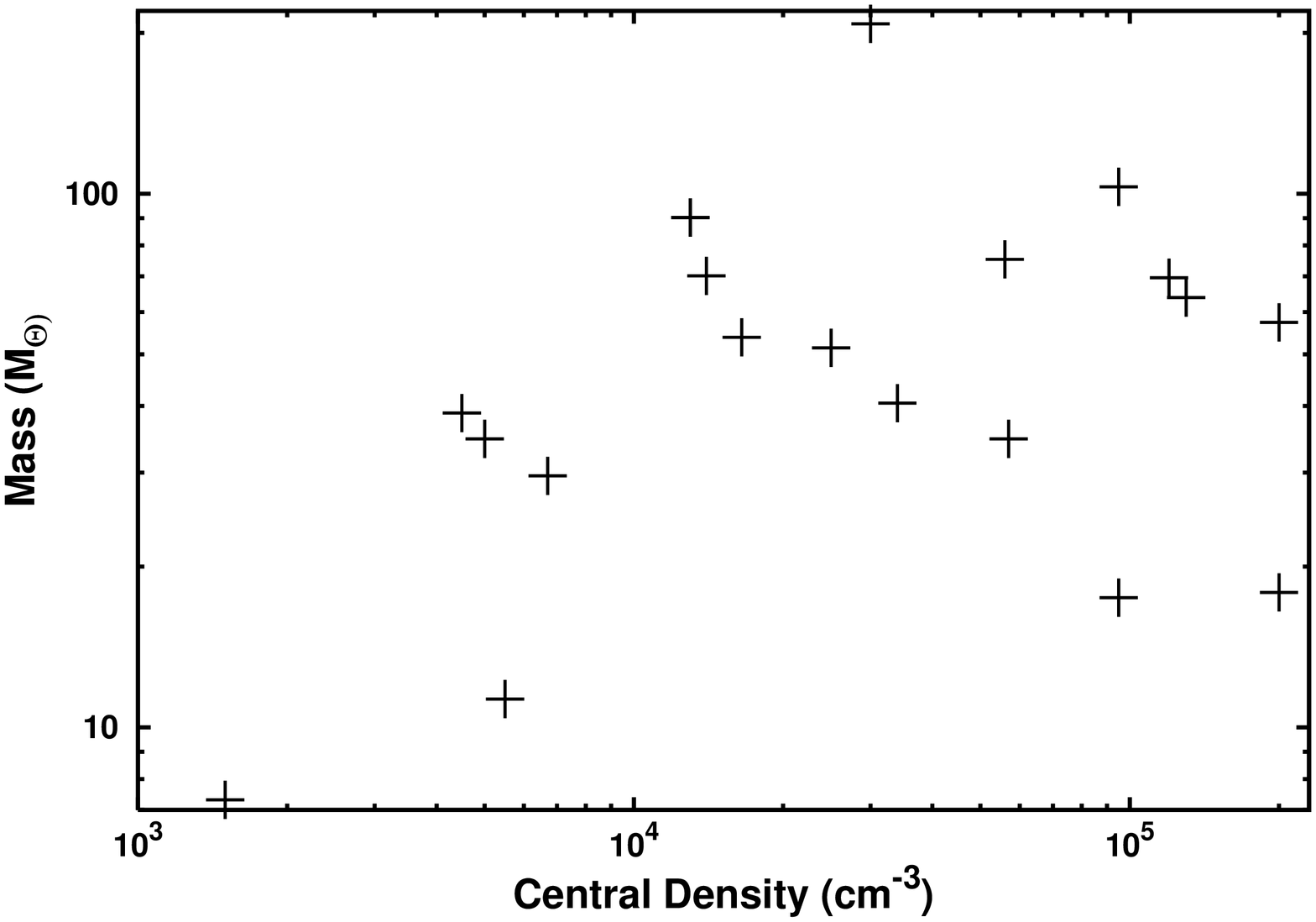}
\caption{Core mass versus radius (left) and central density (right).
No correlation is seen in either case.}
\label{mass_rad_cen}
\end{figure*}

It should be noted that the internal structure of cores within infrared 
dark clouds have not yet been observed in detail. 
Hence, our assumption of an elliptical geometry may be an over-simplification.
If there is structure on smaller scales than we can 
resolve, this might affect our results. For example, small scale fragmentation
might allow the ISRF to penetrate further into the 
cores, meaning that our calculated ISRF values may be upper limits.

\begin{table}
\begin{center}{\small
\caption{The eccentricities of the cores (to 1 decimal place). Column 2 gives the eccentricity when measured from the data and 
column 3 shows 
the eccentricity as measured from the model. Column 4 is the difference between the data and model eccentricities as a percentage 
of the observed eccentricity to 2 significant figures.}\vspace{2mm}
\label{ecc} 
\begin{tabular}{cccc} \hline
IRDC Name& \multicolumn{2}{c}{Eccentricity} & Percentage \\
 & Observed & Model & Difference (\%)\\ \hline
305.798$-$0.097 & 0.9 & 0.9 & 0.0\\ 
307.495+0.660 & 0.8 & 0.9 & 13\\ 
309.079$-$0.208 & 0.8 & 0.9 & 13\\ 
309.111$-$0.298 & 0.9 & 0.9 & 0.0\\ 
310.297+0.705 & 0.5 & 0.6 & 20\\ 
314.701+0.183 & 0.8 & 0.9 & 13\\ 
318.573+0.642 & 0.8 & 0.9 & 13\\ 
318.802+0.416 & 0.8 & 0.9 & 13\\ 
318.916$-$0.284 & 0.8 & 0.9 & 13\\ 
321.678+0.965 & 0.9 & 0.9 & 0.0\\ 
321.753+0.669 & 0.9 & 0.9 & 0.0\\ 
322.334+0.561 & 0.8 & 0.9 & 13\\ 
322.666$-$0.588 & 0.1 & 0.1 & 0.0\\ 
322.914+0.321 & 0.6 & 0.5 & 17\\ 
326.495+0.581 & 0.9 & 0.9 & 0.0\\ 
326.620$-$0.143 & 0.9 & 0.9 & 0.0\\ 
326.632+0.951 & 0.6 & 0.9 & 50\\ 
326.811+0.656 & 0.7 & 0.7 & 0.0\\ 
328.432$-$0.522 & 0.8 & 0.9 & 13\\ 
329.403$-$0.736 & 0.6 & 0.6 & 0.0\\ \hline
\end{tabular}}
\end{center}
\end{table}

\begin{table}
\begin{center}{\small
\caption{The FWHM of the cores at 250\,$\mu$m. Column 2 gives the FWHM when measured from the data and column 3 shows 
the FWHM as measured from the model. Both have errors of $\pm$15\,per\,cent. See text for details.}\vspace{2mm}
\label{fwhm} 
\begin{tabular}{ccc} \hline
IRDC Name& \multicolumn{2}{c}{FWHM at 250\,$\mu$m (\arcsec)} \\
 & Observed & Model \\ \hline
305.798$-$0.097 & 62 & 82\\ 
307.495+0.660 & 55 & 59\\ 
309.079$-$0.208 & 57 & 65\\ 
309.111$-$0.298 & 38 & 43\\ 
310.297+0.705 & 45 & 38\\ 
314.701+0.183 & 40 & 45\\ 
318.573+0.642 & 40 & 44\\ 
318.802+0.416 & 45 & 45\\ 
318.916$-$0.284 & 38 & 42\\ 
321.678+0.965 & 65 & 72\\ 
321.753+0.669 & 85 & 79\\ 
322.334+0.561 & 55 & 63\\ 
322.666$-$0.588 & 42 & 40\\ 
322.914+0.321 & 38 & 42\\ 
326.495+0.581 & 33 & 39\\ 
326.620$-$0.143 & 73 & 73\\ 
326.632+0.951 & 24 & 31\\ 
326.811+0.656 & 43 & 43\\ 
328.432$-$0.522 & 58 & 57\\ 
329.403$-$0.736 & 47 & 51\\ 
\hline
\end{tabular}}
\end{center}
\end{table}

The figures in Appendix \ref{egcores} show the output of the model at 
wavelengths corresponding to those of the observed data. The modelled images have 
pixels of 0.02$\times$0.02\,pc in size and show an area 2.56$\times$2.56\,pc in total. 
All of the model images have been convolved with the appropriate telescope beam size : 2\,\arcsec at 8\,$\mu$m; 
5.2\,\arcsec at 70\,$\mu$m; 11.5\,\arcsec at 160\,$\mu$m; 18\,\arcsec at 250\,$\mu$m; 25\,\arcsec at 350\,$\mu$m;  and 
36\,\arcsec at 500\,$\mu$m \citep{spitzer, spire}. 
For wavelengths where no emission is visible in the model, an image showing 
background radiation is shown. Note that no attempt was made 
to correctly model the surrounding area, so these images are not an accurate 
representation of the parent cloud but do show the cores 
in absorption as seen in the observations -- see also 
\citet{stamatellos10} and \citet{wilcock11}.

\section{Results}\label{results}

\subsection{Masses}\label{mass}

\begin{figure}
\includegraphics[angle=-90,width=85mm]{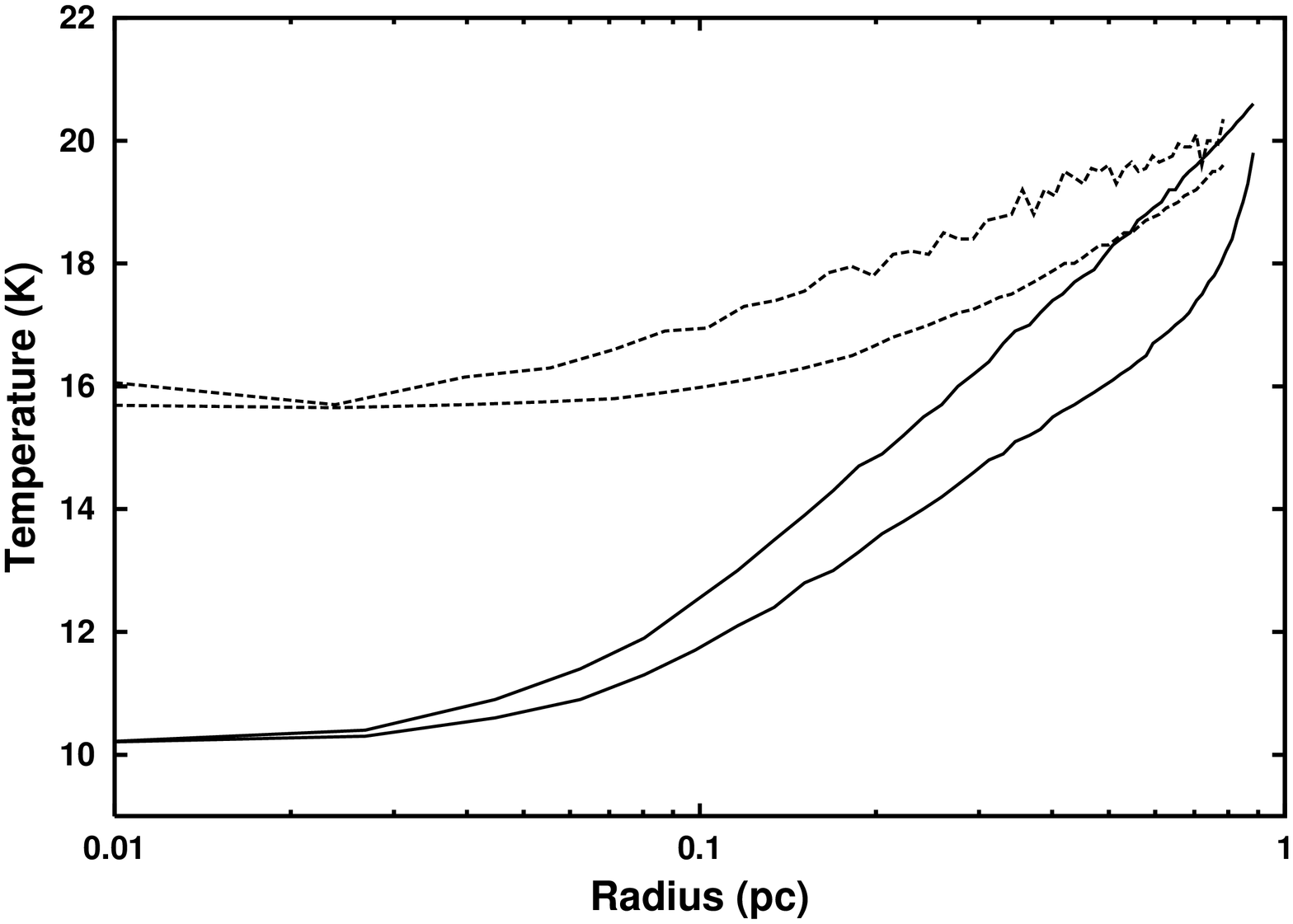}
\caption{Temperature profiles of two typical cores. The solid line shows 
core 310.297+0.705 and the dashed line shows core 321.753+0.669.
The top line of each pair shows a direction of $\theta$=0\,\degr, 
perpendicular to the midplane of the core, and the lower line 
shows $\theta$=90\,\degr, along the midplane of the core.} 
\label{temp_pro}
\end{figure}

\begin{figure}
\includegraphics[angle=-90,width=85mm]{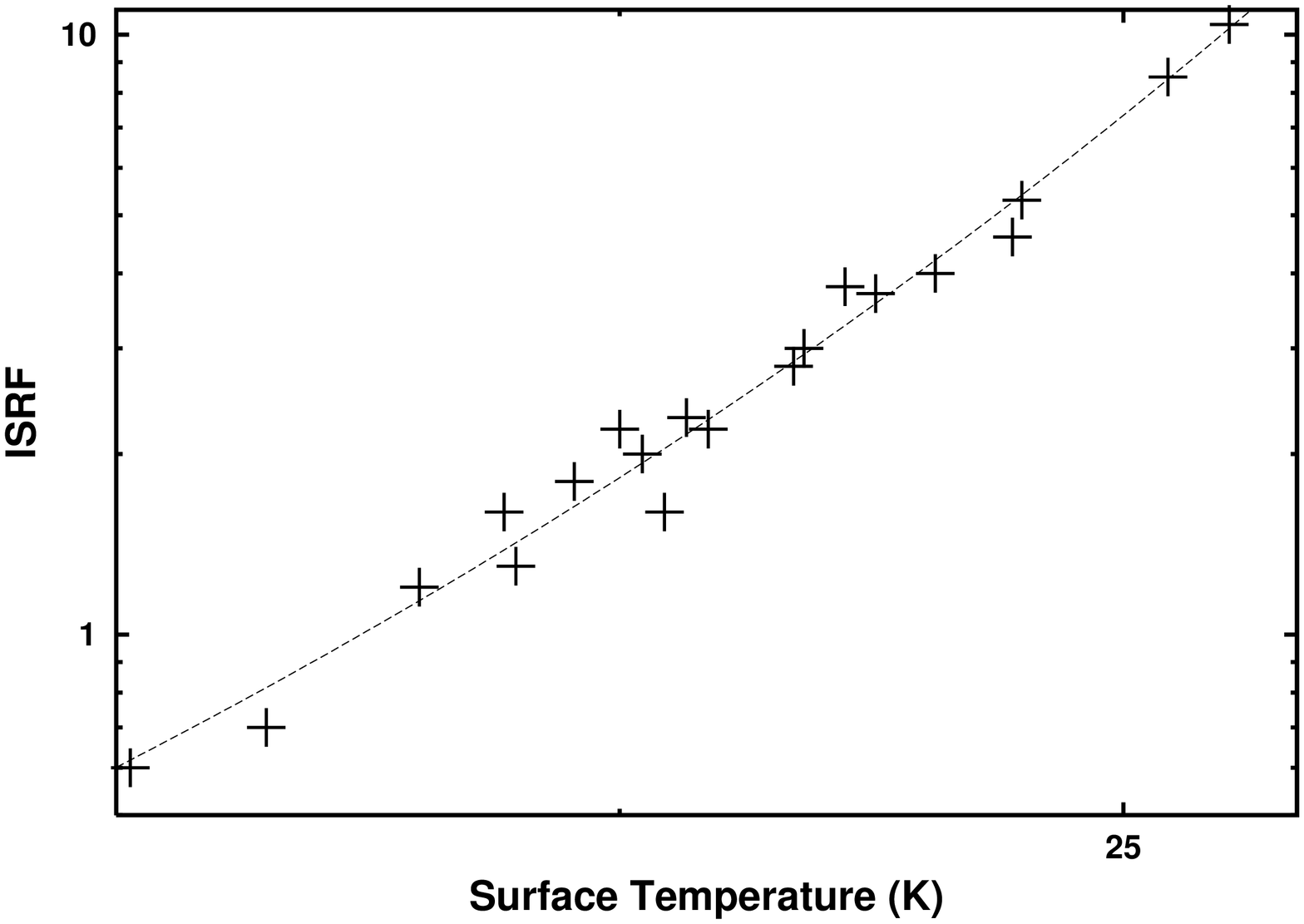}
\includegraphics[angle=-90,width=85mm]{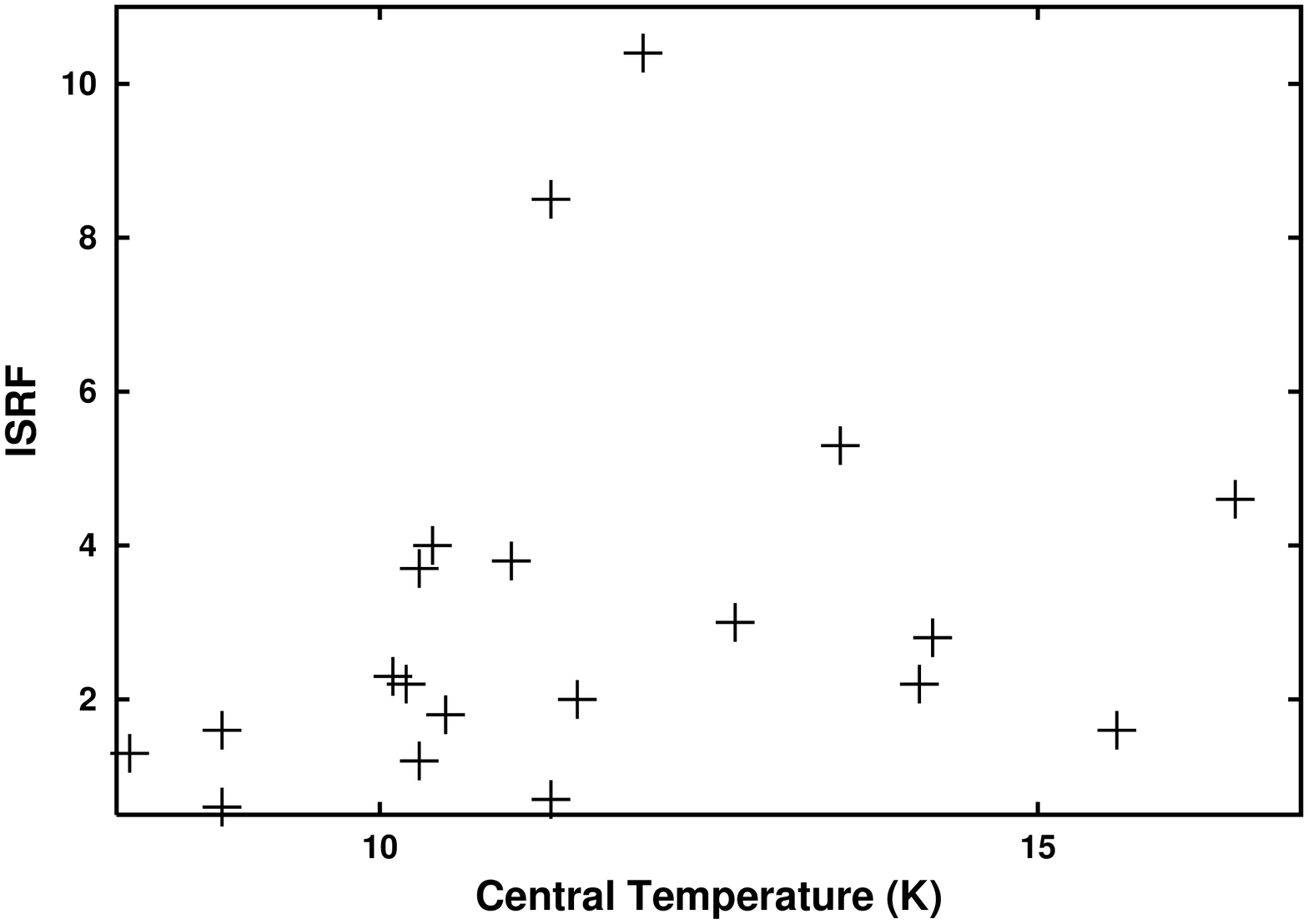}
\caption{Upper: The ISRF in multiples of the \protect\citet{black94} 
radiation field plotted against the surface temperature. 
The dashed line shows the relation between the surface
temperature and ISRF as described in Equation \ref{6:eqnmax_isrf}. 
Lower: The ISRF in multiples of the \protect\citet{black94} 
radiation field plotted against the central temperature.}
\label{temp_isrf}
\end{figure}

\begin{figure}
\includegraphics[angle=-90,width=87mm]{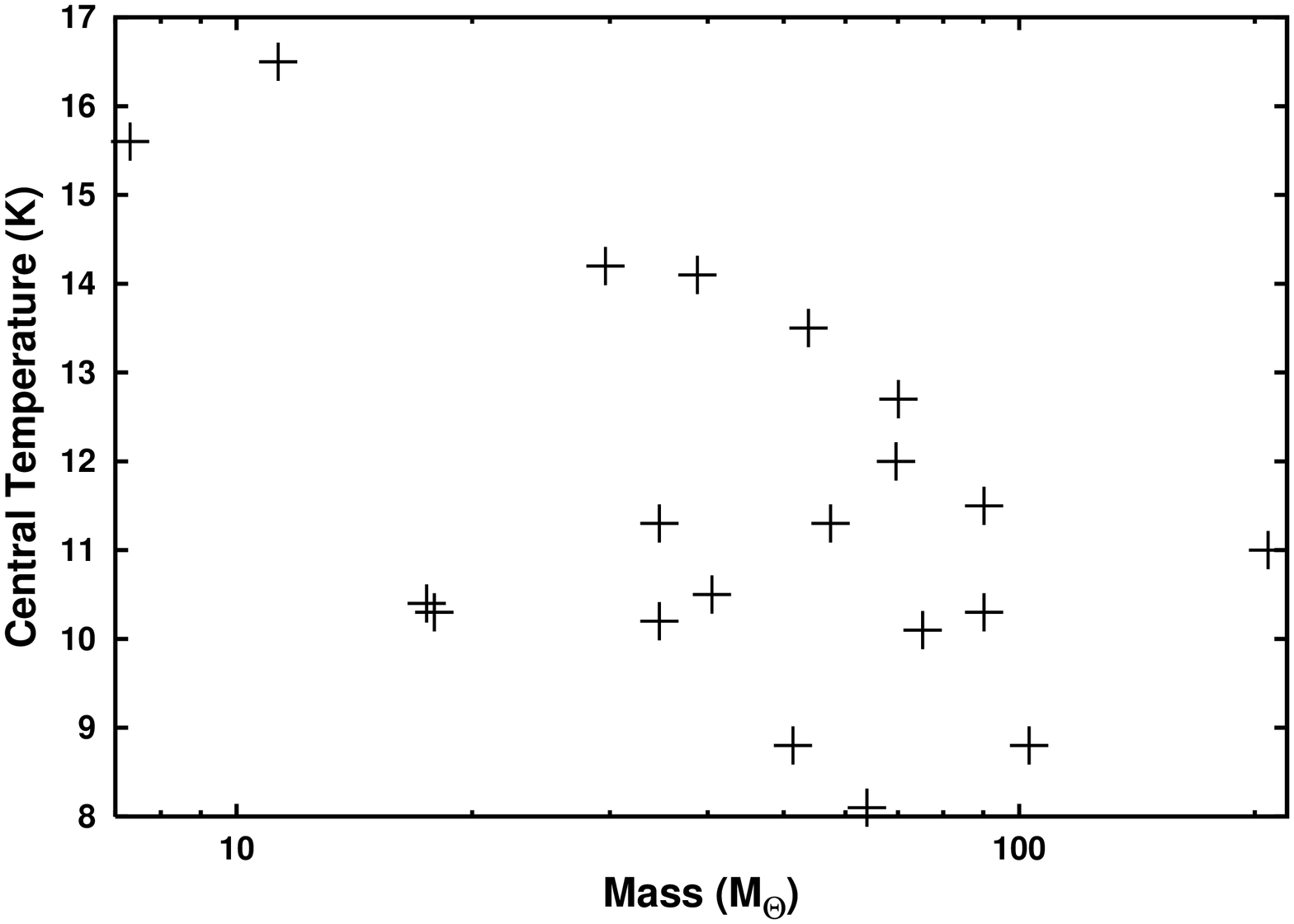}
\includegraphics[angle=-90,width=87mm]{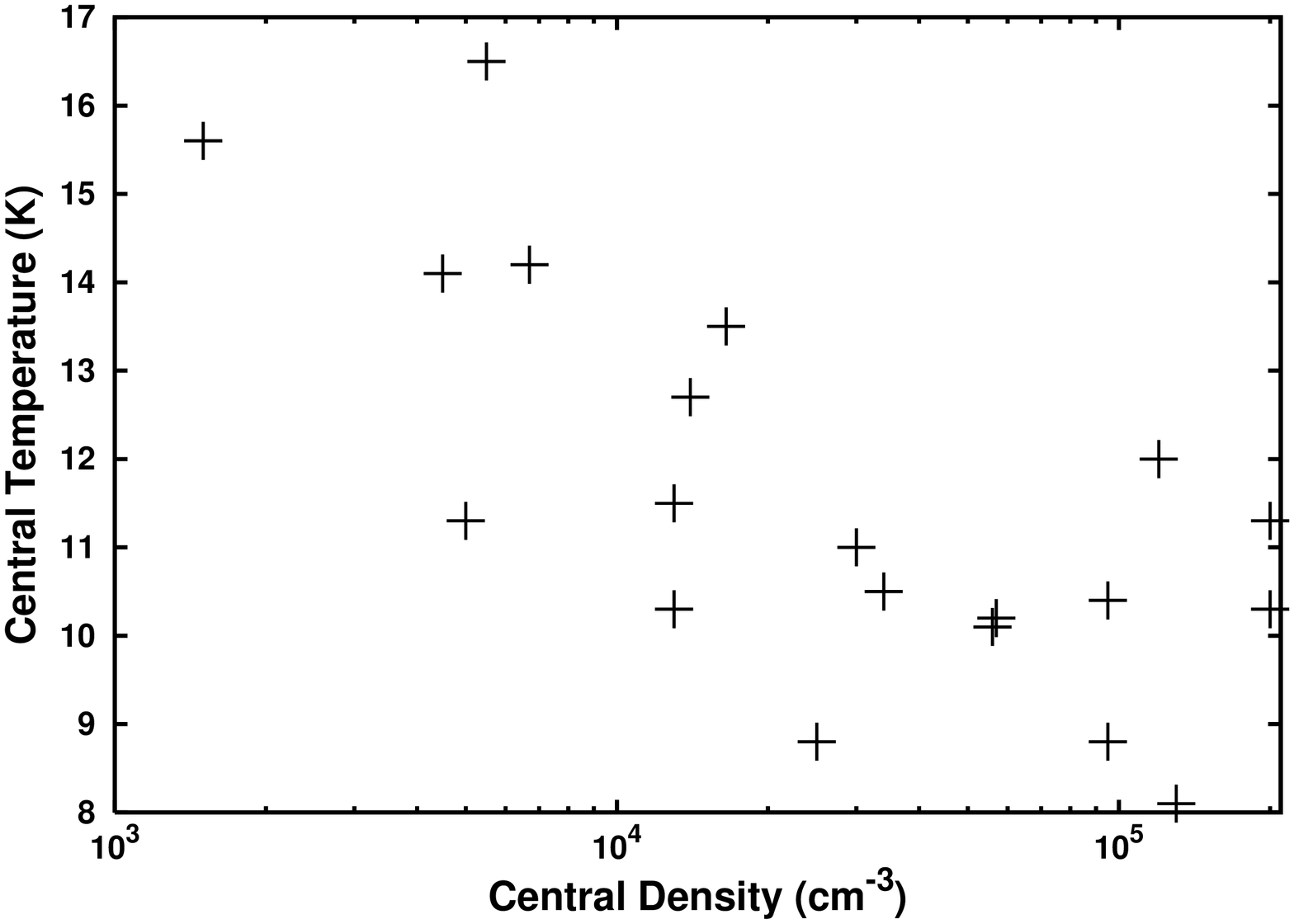}
\caption{Upper: The mass of the cores plotted against their central 
temperatures. Lower: 
The central density of the cores plotted against their central temperatures.} 
\label{min_cen}
\end{figure}

The core masses were calculated within the \textsc{Phaethon} code using an 
opacity of 
$\kappa_{500\,\mu m}$=0.03\,cm$^{2}$\,g$^{-1}$ \citep{ossenkopf94}. Our 20 
cores have masses in the range 
$\sim$7-200\,M$_{\odot}$, with a mean of 58\,M$_{\odot}$ and a median of 
52\,M$_{\odot}$. Their distribution is shown in Figure~\ref{hist}.

\citet{rathborne06} calculated the masses of 190 infrared dark 
cores and found a range between 10 and 2100\,M$_{\odot}$, with a 
median of 120\,M$_{\odot}$. However, they also find that 67\,per\,cent of 
their cores have a mass between 30 and 300\,M$_{\odot}$. 
By choosing only a few of the youngest, most nearby
cores, we are more likely to only pick 
those with lower masses, so this discrepancy is most likely a selection effect.
Figure~\ref{mass_rad_cen} shows the masses of our cores plotted 
against their radii and central densities. The masses of the cores 
do not appear to be correlated with either parameter.
%The relationship between radius and central density is discussed in Section \ref{cenden}.

\subsection{Temperatures}\label{temp}

When selecting our sources the aim was to pick the youngest 
subset of infrared dark cores: those which show no internal MIR source 
and which were thus deemed to be starless. The lack of a MIR 
source makes it unlikely that these cores have any significant heating
from within.
The cores are, instead, heated by external radiation impinging onto their 
surfaces. 
This external heating results in a temperature gradient throughout 
each core, decreasing from surface to centre. Temperatures in individual 
cores have been observed to vary by up to 15\,K 
\citep{peretto10,wilcock11}. It should be noted that the temperature 
profiles of the cores are not linear but rather show an
exponential relationship with distance from the centre of the 
core. Two example temperature profiles are
shown in Figure~\ref{temp_pro}.

In Table \ref{coreprop}, the temperature at the centre of the 
core and at the surface are given. These have
averages of 11 and 21~K respectively. The dust temperatures in flattened cores show
a variation at different directions within the cores. The cores are colder in their
midplanes ($\theta=90$\,\degr) than in the direction perpendicular to their
midplanes ($\theta=0$\,\degr). The outer layers of the cores show the biggest
variation with direction. The surface temperatures show a greater
range (varying by over 10\,K) than the central temperatures (which have a range of less than 7\,K). 
Figure \ref{hist} shows the distribution of both sets of temperatures.

\subsubsection{Surface Temperatures}

Figure \ref{temp_isrf} shows the ISRF surrounding each core plotted 
against both the surface (upper panel; on a log scale) and central (lower panel) 
temperatures. The surface temperature is
seen to be strongly dependent on the 
amount of radiation falling on the core. As a core's major
source of heating is from external radiation, this is as expected. 
The higher the local ISRF, the more radiation will fall on
the core and the higher its surface temperature will be. The two show an 
exponential relationship, with increasing amounts of radiation 
needed to heat the core to greater temperatures. 

The correlation 
coefficient between the ISRF and T$_s$ is r=0.99, p$<$0.05. r is the Pearson-product 
moment correlation coefficient and p is the level of significance (i.e. p$<$0.05 implies
a 95\,per\,cent confidence that a relationship exists).

The dotted line in Figure \ref{temp_isrf} shows the best-fit canonical relationship: 
\begin{equation}\label{6:eqnmax_isrf}
%\textrm{ISRF}=\left(0.007\pm0.001\right)e^{ \left(\textrm{0.278} \pm \textrm{0.007}\right) \textrm{T}_s }, 
\ln \left( \textrm{ISRF} \right) = 5.8 \pm 0.2 \ln \left( \textrm{T}_s \right) - 16.8 \pm 0.6
\end{equation}
where ISRF is the radiation field found in the vicinity of the core in multiples of the 
\citet{black94} radiation field and T$_s$ is the core's surface temperature.

%From Equation \ref{eqnmax_isrf}, we extrapolate that for a core to have a 
%surface temperature over 30\,K the ISRF would need to be 
%over 30 times that of the solar neighbourhood and for a temperature over 
%35\,K the ISRF would have to be 100 times greater. Most 
%infrared dark cores will therefore have a temperature lower than 30\,K. 
%The majority of our cores have temperatures less than 25\,K. 

%Similarly, a core with a surface temperature of 10\,K would need to be 
%situated in a radiation field of only one-tenth of the solar 
%value. This makes lower temperatures very unlikely, albeit not impossible 
%if, for example, the core was buried very deeply within 
%the parent cloud. All of our cores had surface temperatures greater than 16\,K.

From Equation \ref{6:eqnmax_isrf}, we extrapolate that for a core to have a 
surface temperature over 30\,K the ISRF would need to be over 20 times that of the solar neighbourhood 
and for a surface temperature over 
40\,K the ISRF would have to be 100 times greater. Most infrared dark cores will therefore have a surface 
temperature lower than 30\,K. The majority of our cores have surface temperatures less than 25\,K. 

Similarly, a core with a surface temperature less than 10\,K would need to be situated in a radiation field 
of only one-hundredth of the solar value. This makes lower temperatures very unlikely, although not 
impossible. All of our cores had surface temperatures greater than 16\,K.

\begin{figure}
\includegraphics[angle=-90,width=85mm]{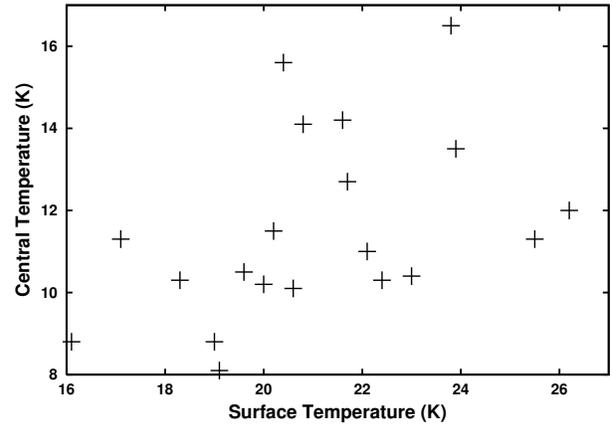}
\caption{The temperature at the centre of the core versus
the temperature at the surface of the core.} 
\label{min_max}
\end{figure}

\begin{figure}
\includegraphics[angle=-90,width=85mm]{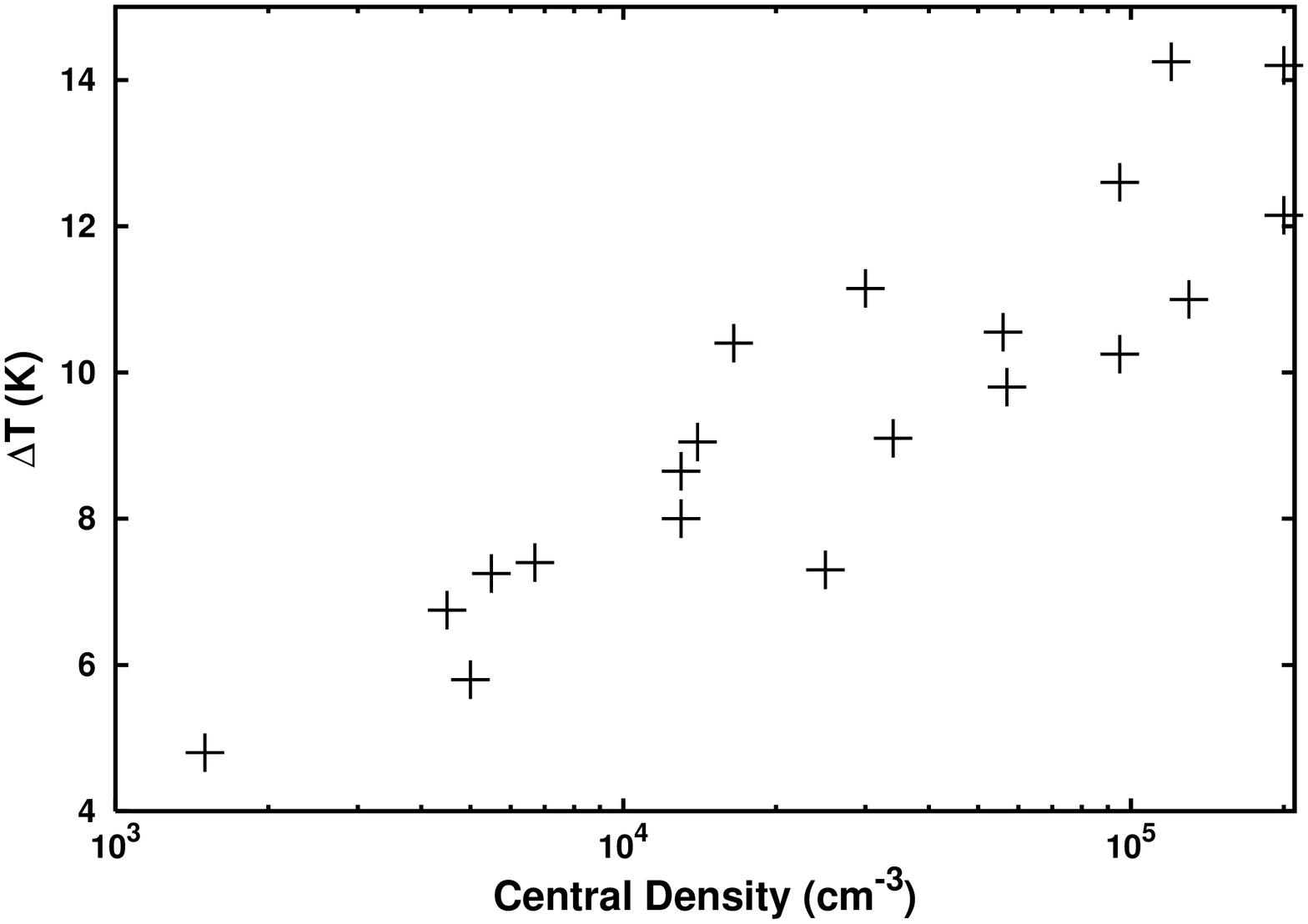}
\includegraphics[angle=-90,width=85mm]{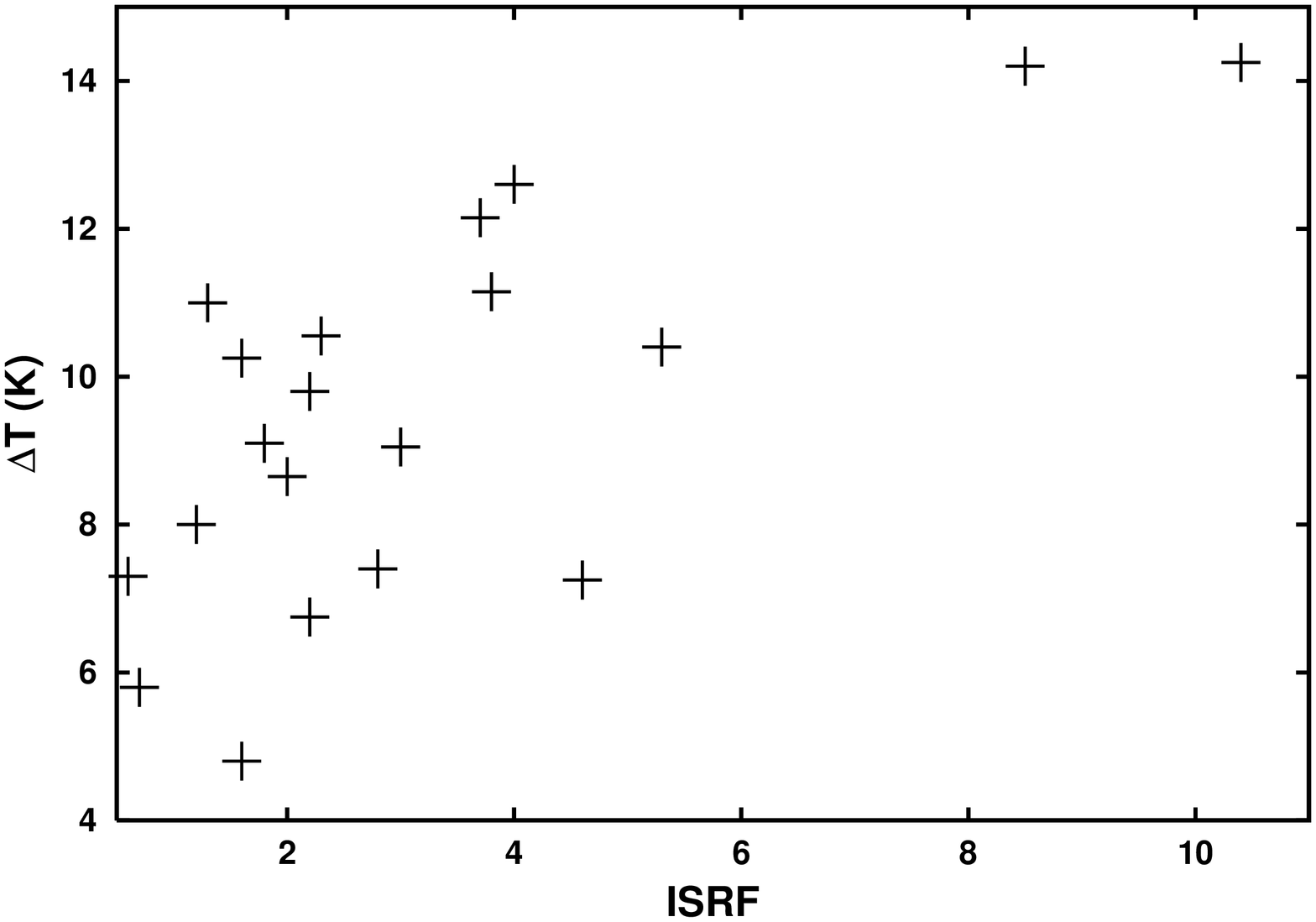}
\includegraphics[angle=-90,width=85mm]{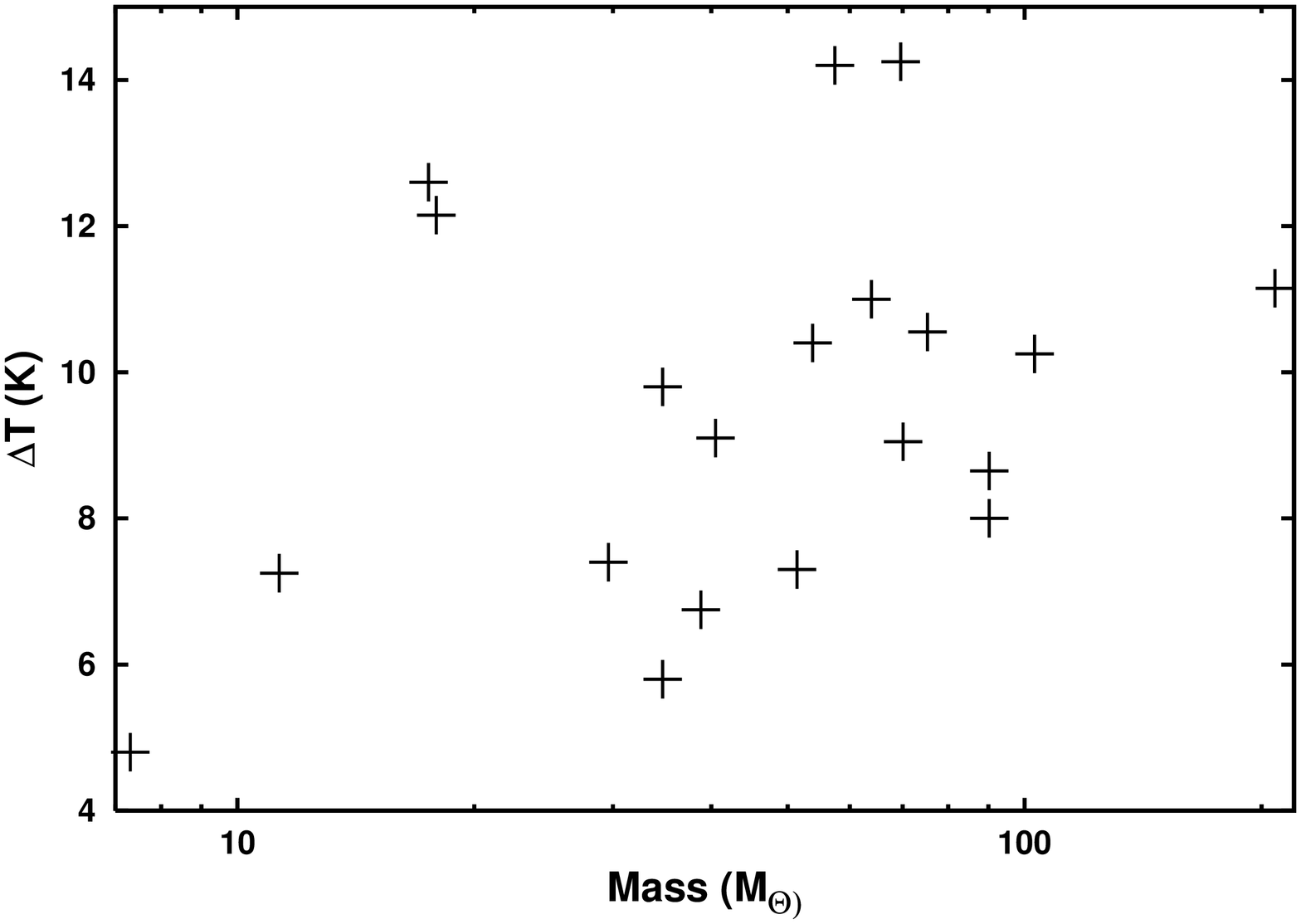}
\caption{Upper: The central density of the cores plotted against 
the difference between the surface and central 
temperatures ($\Delta$T). Middle: The ISRF in multiples of the 
\protect\citet{black94} radiation field plotted against $\Delta$T. 
Lower: The masses of the cores plotted against $\Delta$T.} 
\label{tdif}
\end{figure}

\subsubsection{Central Temperatures}

The temperatures at the centres of the cores vary according to different 
criteria to those at the surface. 
The central temperatures of the cores do not show any dependency on 
the ISRF -- see the lower panel of Figure \ref{temp_isrf}. 
Instead, the temperature at the centre of the core appears to be 
determined by its mass and central density, as shown in 
Figure~\ref{min_cen}. 

The central temperature shows an inverse linear relationship with 
central density with a correlation coefficient of r=-0.46, p$<$0.05. 
As radiation falls onto the core surface
it will heat the outer layers first, giving the temperature 
gradient visible in all IRDC cores. In denser cores the 
radiation will have a greater probability of being absorbed in 
the outer layers and less will travel through the core to
heat the centre. Thus, denser cores have a lower central 
temperature than cores with a lower density. 

The masses of the cores show no relation to the temperature at the 
surface of the core but do
show a weak anti-correlation with the central temperatures, r=-0.35, p$<$0.05.
Cores with high masses have a lower central temperature. This is due 
to the same effect that causes the central temperature to vary
with density. 
As mentioned in Section \ref{mass}, mass and central density show 
no correlation with one another. This implies that their 
relationships with central temperature must be independent of each other.

Figure~\ref{min_max} shows the central temperature 
plotted against the surface temperature. The two show a broad 
correlation (r=0.44, p$<$0.05), implying that cores with a similar central density, 
have a higher central temperature if there is a greater ISRF. 
However, the cores with highest surface temperatures do not 
necessarily have the highest central temperatures. For example, 
cores 307.495+0.660 and 326.811+0.656 have vastly different surface 
temperatures but identical central temperatures.

\subsubsection{Temperature Range}

The range of temperatures in each core, $\Delta$T, is found via
\begin{equation}
\Delta$T$=$T$_s-$T$_c,
\end{equation}
where T$_c$ is the central temperature of the core, and 
T$_s$ is the surface temperature of the core.
$\Delta$T is indicative of a core's temperature profile. A core with a high
$\Delta$T will show a steeper profile than a core with a 
lower $\Delta$T. Figure \ref{temp_pro} shows the temperature
profiles of two cores with similar radii. G310.297+0.705 
has $\Delta$T$=10.5$\,K and a far steeper profile than G321.753+0.669
which has $\Delta$T$=4.8$\,K.
$\Delta$T appears to be dependent on a core's central density (r=0.81, p$<$0.05)
and the surrounding ISRF, but not with its mass (r=0.27, p$>$0.05). This is shown 
in Figure~\ref{tdif}. 
A possible exponential relation between $\Delta$T and the ISRF can 
be seen. As this is similar to the relationship between ISRF 
and the surface temperature, it is likely a result of the $\Delta$T 
dependence on surface temperature of the core. 

The strongest relation is that between $\Delta$T and the core 
central density. The difference 
between the temperature at the surface and centre of a
core clearly increases as the central density increases. Radiation 
passing through denser material has a greater 
probability of being absorbed in the surface layers of the core, and 
so less radiation passes through to (and heats) the centre of 
the core. Therefore, if two cores are situated within areas of 
equivalent ISRFs then, while the surface temperatures would be the 
same, the denser core would have a colder centre and exhibit a 
wider range of temperature.

\begin{figure}
\includegraphics[angle=-90,width=85mm]{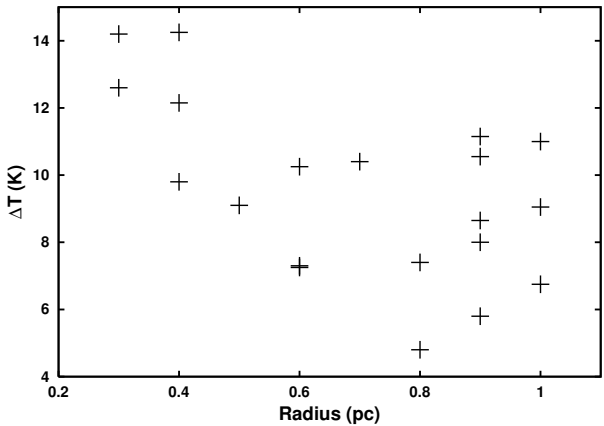}
\caption{$\Delta$T versus radius.} 
\label{rad_tdif}
\end{figure}

$\Delta$T also shows a weak inverse correlation with the radius 
of a core (r=-0.55, p$<$0.05) -- see Figure~\ref{rad_tdif}. The larger the size of the 
core, the smaller the temperature difference between the surface 
and the centre. This is the opposite of what one would generally
expect and, as it is a weak relationship, is not likely to be a direct 
result but rather a side effect of the fact that our smallest cores have the 
highest densities.

\subsection{Central Densities} \label{cenden}

The central densities of the cores vary from 
1.5$\times10^3$\,cm$^{-3}$ to 2.0$\times10^5$\,cm$^{-3}$, with a mean of 
2.0$\times10^4$\,cm$^{-3}$. The central density of the 
core is an input to the \textsc{Phaethon} code and mostly affects the longer
wavelength end of the SED. Figure~\ref{hist} shows the 
distribution of central densities.
The density profile of each core is calculated via 
Equation~\ref{densityeqn}. This results in the density at the surface of 
the core (where it is at its lowest) directly correlating 
with the maximum, central density. 
It is simply an input definition of the model.
%This is shown in Figure~\ref{den_den}. 
The surface densities range from 0.1 to 20\,cm$^{-3}$.

%\begin{figure}
%\includegraphics[angle=-90,width=87mm]{fig11.eps}
%\caption{Central versus surface density.} 
%\label{den_den}
%\end{figure}

\begin{figure}
\includegraphics[angle=-90,width=87mm]{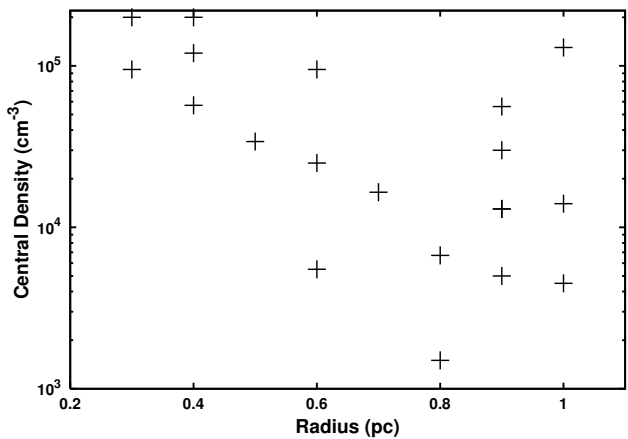}
\caption{Central density versus radius.} 
\label{rad_cen}
\end{figure}

Figure~\ref{rad_cen} shows the central density of each core plotted
against its radius. There is no obvious correlation between
the two but there is a noticeable lack of cores in the lower 
left-hand corner with both a low 
central density and a small radius. 
This is likely to be a selection effect. 
This implies that: either small, low density 
cores do not show up as a distinct peak against their parent cloud;
or they do not block enough MIR background emission to show a 
significant dip at 8\,$\mu$m; 
or that they are simply fainter and so are more likely to be 
confused at 500\,$\mu$m. 

This latter theory was tested by creating two model cores that would reside in 
the lower left of Figure~\ref{rad_cen}. Both had a radius of
0.3\,pc, equal to the smallest core we model. One had a central 
density of $2\times10^3\,\textrm{cm}^{-3}$ and the other had
$2\times10^4\,\textrm{cm}^{-3}$. 
We placed them both at a distance of 3.1\,kpc, gave them an 
ISRF of 3.2 times the \citet{black94} radiation field and an asymmetry 
factor of 2.5 (these are the average parameters of our 
modelled cores). The resulting objects have masses of 0.4\,M$_{\odot}$ 
and 3.9\,M$_{\odot}$. The cores were added to the Hi-GAL data in several regions
and studied as with the original PF09 cores. Even in very low background regions,
neither core met our criteria for isolation at 500\,$\mu$m. They would 
not have been selected by our observations.

\subsection{The Interstellar Radiation Field}

The ISRF for each core is quoted in terms of multiples of the 
\citet{black94} radiation field. The \citet{black94} radiation field
models the ISRF in the solar neighbourhood and measures a value of
$1.3\times10^{-16}$\,erg\,s$^{-1}$\,cm$^{-2}$\,Hz$^{-1}$\,sr$^{-1}$ at 
500\,$\mu$m. Along with the central density, the ISRF is one of the 
inputs of the \textsc{Phaethon} code (see Section~\ref{modelling}), 
and mostly affects the SED at shorter wavelengths. 
The ISRF around our cores ranges from 0.6 to 10.4 times 
the \citet{black94} field, with an average multiple of 3.2. 
Figure~\ref{hist} shows the distribution of the cores' ISRF.

The amount of radiation incident on our cores is 
typically greater than that which we 
observe in the solar neighbourhood. This is 
consistent with the inner 
Galactic Plane generally having a stronger radiation field
than that of the solar neighbourhood. 
However, two of the cores do show a radiation 
field lower than the solar value. These 
are 307.495+0.660 and 321.678+0.965 (see Table~\ref{coreprop}). 
There are two possible explanations for this: either they 
simply exist in an area which has a particularly low ISRF; 
or the ISRF surrounding the parent IRDC is consistent with other 
cores but the parent cloud is absorbing most of the radiation. 
The latter could be caused by the IRDC being especially dense, or by 
the cores being more deeply embedded within the parent cloud. 
Either case would make it difficult for radiation to penetrate 
far enough to affect the cores.

\begin{figure}
\includegraphics[angle=-90,width=87mm]{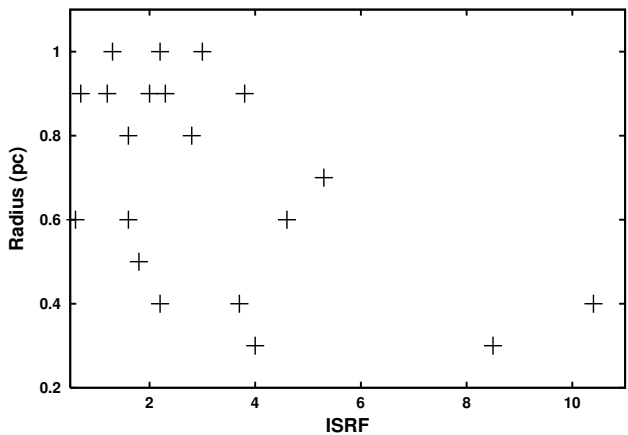}
\caption{Core radius versus ISRF -- in multiples 
of the \protect\citet{black94} radiation field.} 
\label{rad_isrf}
\end{figure}

Figure \ref{rad_isrf} shows a plot of core radius versus ISRF. 
We detect cores of all sizes with ISRFs at the lower
end of our scale but our modelled sample doesn't include any 
cores with a high ISRF and a large radius.
If radiation pressure were forcing the cores to condense 
then we would expect the central density to increase as the radius 
decreased. While our smaller cores do tend to have densities at the 
higher end of our range, the larger cores have both low and high densities.
This is due to the selection effect discussed in Section~\ref{cenden}. 
We therefore dismiss radiation pressure as the reason why we
detect no large cores in areas of high ISRF.
Instead, we consider that in areas with a high radiation field the 
outer layers of large cores may `boil away' due to the large amounts
of energy being absorbed. This would explain why we only see 
small cores in areas of high ISRF and why there is no significant
increase in the density of these cores.

The radius and surface temperature appear to show a similar, 
albeit inverse, relationship to that of ISRF and radius. 
We speculate that the radius of a core is actually affected by 
the ISRF and that any correlation between temperature and radius 
exists as a result of the dependence of 
the surface temperature on the ISRF. 

\subsection{Asymmetry Factor}\label{asym}

The asymmetry factor input to the model controls the shape of the 
core. A value of 1.0 would result in a circular core. The
ellipticity of the core is increased by raising the asymmetry factor 
up to a maximum of 3.0. Most IRDC cores are 
elongated and have a high eccentricity. 80\,per\,cent of our cores have an 
asymmetry factor greater than 2 -- see Table~\ref{coreprop}. It is also
possible that some of our more circular cores could be elongated but viewed 
end-on. 

%None of the other physical properties appear to have any correlation 
%with the ellipticity of the core. Elongated cores show similar
%masses, temperatures and densities to the more circular cores. For 
%example, 307.495+0.660 and 326.495+0.581 are both
%elongated (with asymmetry factors of 3.0 and 2.9, respectively) but 
%have vastly different radii, ISRFs and densities -- see Table~\ref{coreprop}.
%Whereas 326.811+0.656 has a much lower eccentricity 
%(with an asymmetry factor of 1.4) but is similar in size, mass 
%and temperature to 326.495+0.581. This implies that the shape of the 
%core has little or no effect on the other physical properties 
%of the core. 
%or on its evolution.

\subsection{Mean Core Properties}
\begin{table}
\begin{center}
\caption{The mean values of the physical properties of the 20 modelled cores.}
\label{mean}
\begin{tabular}{lc} \hline
Physical Property & Mean \\ \hline 
Radius (pc) & 0.7 \\
Central Density (cm$^{-3}$) & 5.6$\times$10$^4$ \\ 
Mass (M$_{\odot}$) & 58.3\\
ISRF & 3.2\\
Central Temperature (K) & 11.5 \\
Surface Temperature (K) & 21.0 \\ 
$\Delta$T (K) & 9.5 \\
Asymmetry Factor & 2.5\\
\hline
\end{tabular}
\end{center}
\end{table}

Table \ref{mean} shows the mean values of each of the eight physical properties 
discussed in Sections \ref{mass}--\ref{asym}. The mean physical properties reiterate the fact that 
starless cores are dense, elongated, cold objects and, when found in the Galactic Plane, have a local 
ISRF greater than that which is found in the solar neighbourhood. Within the cores, the temperature 
decreases from surface to centre, typically by $\sim$10\,K. The mean asymmetry factor is 2.5, 
corresponding to an aspect ratio of $\sim$1:7.

\section{Summary}

Using data from \textit{Herschel} and \textit{Spitzer}, 20 isolated,
starless cores were modelled with the \textsc{Phaethon} radiative 
transfer code. A
flattened density profile, as described by Equation \ref{densityeqn}, was assumed for all. 
The FWHM of each core was measured at 250\,$\mu$m and used as the core's radius. The
radial distance for which the central density is constant was set to one tenth of this value. 
The density and 
surrounding ISRF were varied until the model SED matched the
observed SED at 160, 250, 350 and 500\,$\mu$m.

Output parameters were measured from the model. The IRDC
cores were found to have masses ranging from around 
ten to around a few hundred 
solar masses. The masses of the cores were found not to correlate 
with their size or central density.
The temperature at the surface of a core was seen to depend
almost entirely on the level of the ISRF surrounding the core. 
No correlation was found between the temperature at the centre of a core 
and its local ISRF. The central temperature was seen to depend, instead, 
on the density and mass of 
the core. A core with a high density or mass is likely to have a 
lower central temperature. 

The range of temperatures from centre to edge in any one core was found to
correlate strongly with the central density of that core. 
Low density cores exhibit a much smaller range of 
temperatures and thus a shallower
temperature gradient than higher density cores. The 
cores situated in a high ISRF (and thus with a high surface 
temperature) are also likely to show a greater range of temperature
from centre to edge. 

No large cores were found within an area of comparatively
high ISRF. However, there appears to be no correlation between the 
ISRF surrounding a core and its central density.
Hence, we rule out radiation 
pressure as the cause. Instead, we consider that the outer layers of 
cores in areas of high radiation may be evaporated due to the high
levels of energy being absorbed, leaving only smaller cores 
in high-ISRF regions.
16 out of 20 isolated IRDC cores
showed high levels of ellipticity. 
%However, there was no difference detected in 
%the mass, radius, density or environment of cores that are highly 
%elliptical compared to those that are circular. This led us to the 
%conclusion that the shape of a core has little effect on its other physical parameters.

\section*{Acknowledgments}
LAW gratefully acknowledges STFC studentship funding.
SPIRE was developed by a consortium of institutes led by Cardiff 
University (UK) and including Univ. 
Lethbridge (Canada); NAOC (China); CEA, LAM (France); IFSI, Univ. Padua 
(Italy); IAC (Spain); 
Stockholm Observatory (Sweden); Imperial College London, RAL, UCL-MSSL, 
UKATC, Univ. Sussex 
(UK); and Caltech, JPL, NHSC, Univ. Colorado (USA). This development has 
been supported by national 
funding agencies: CSA (Canada); NAOC (China); CEA, CNES, CNRS (France); 
ASI (Italy); MCINN 
(Spain); SNSB (Sweden); STFC (UK); and NASA (USA). 
PACS was developed by a consortium of institutes led by MPE (Germany) 
and including UVIE
(Austria); KU Leuven, CSL, IMEC (Belgium); CEA, LAM (France); MPIA 
(Germany); INAF-IFSI/OAA/OAP/OAT, LENS, SISSA (Italy); IAC
(Spain). This development has been supported by the
funding agencies BMVIT (Austria), ESA-PRODEX (Belgium), CEA/CNES (France), 
DLR (Germany),
ASI/INAF (Italy), and CICYT/MCYT (Spain). 
HIPE is a joint development by the \textit{Herschel} Science Ground 
Segment Consortium, consisting of ESA, the NASA \textit{Herschel} Science 
Center, and the HIFI, PACS and 
SPIRE consortia.

\bibliographystyle{mnras.bst}
\bibliography{bib}

\appendix
\section{Examples of the modelled cores}\label{egcores}

\begin{figure*}
\includegraphics[angle=-90,width=55mm]{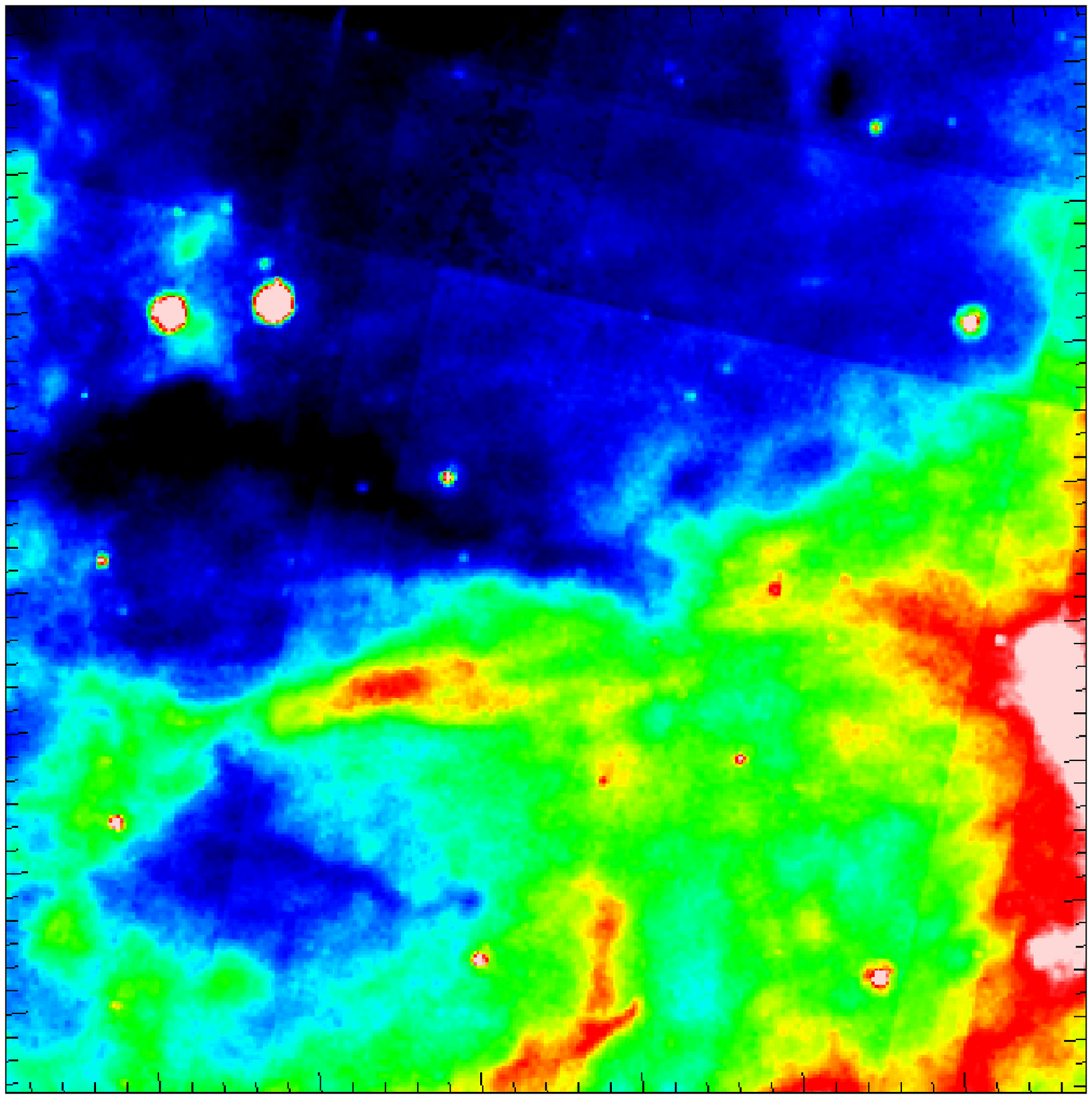}
\includegraphics[angle=-90,width=55mm]{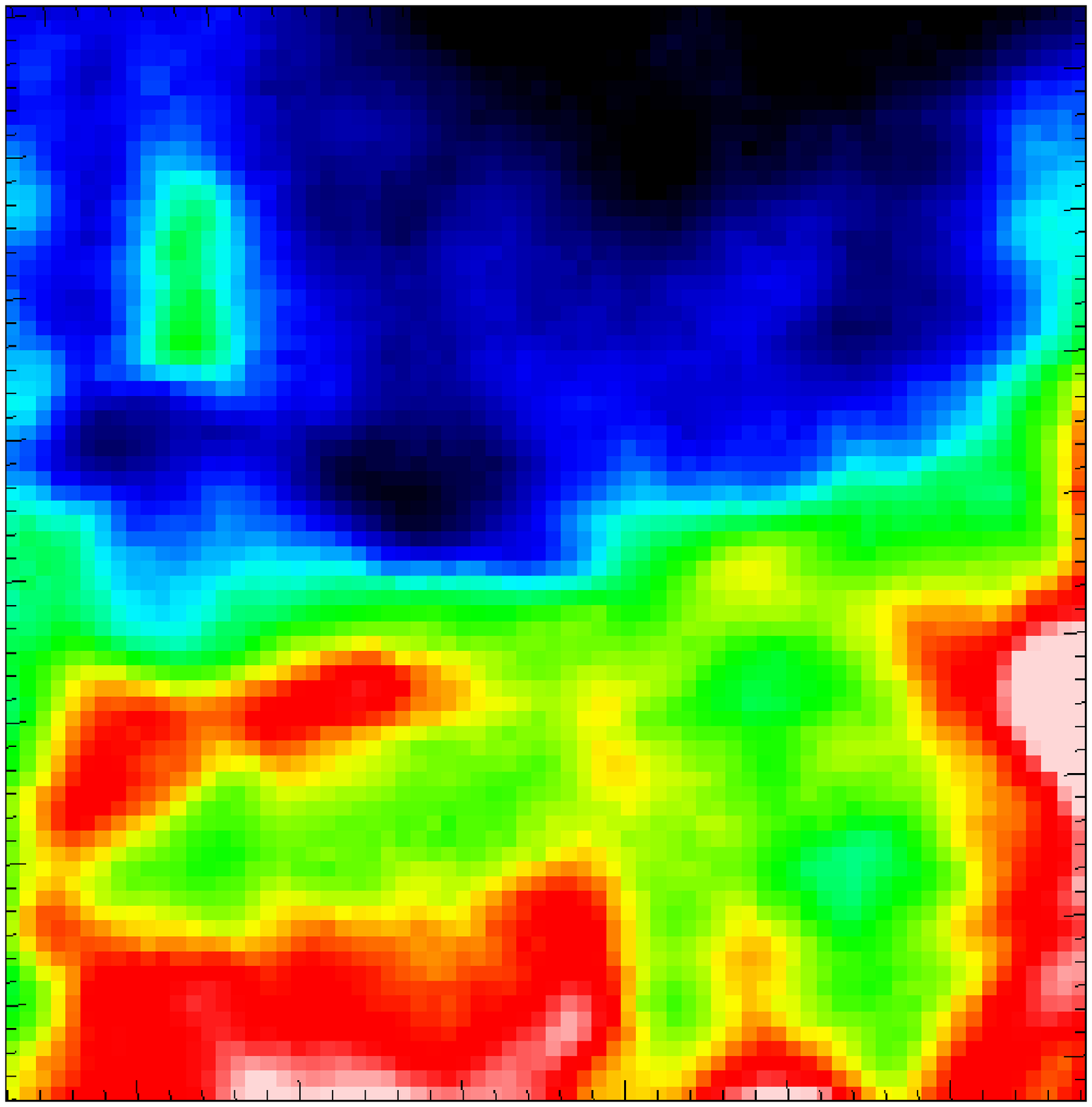}
\includegraphics[angle=-90,width=55mm]{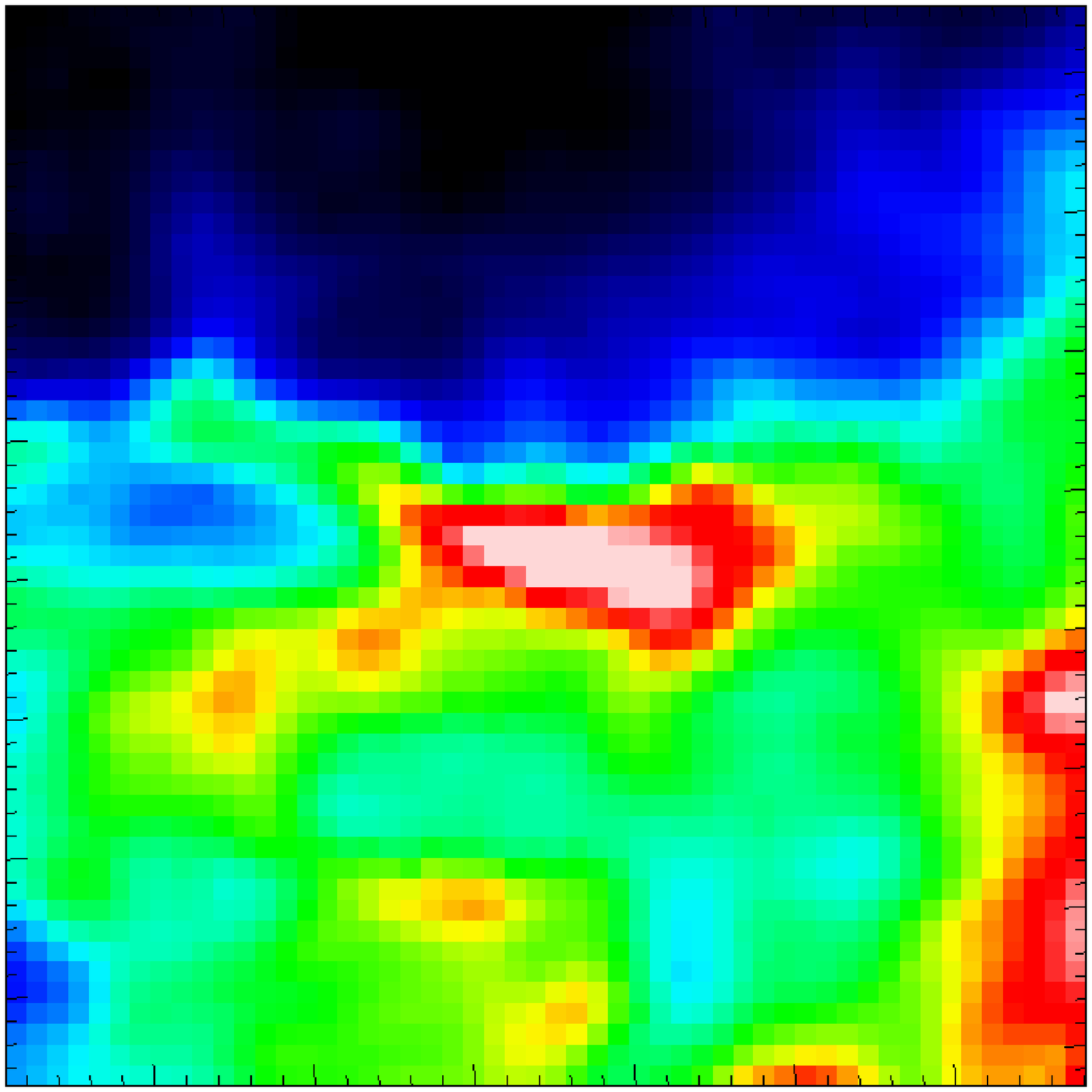}
\includegraphics[angle=0,width=55mm]{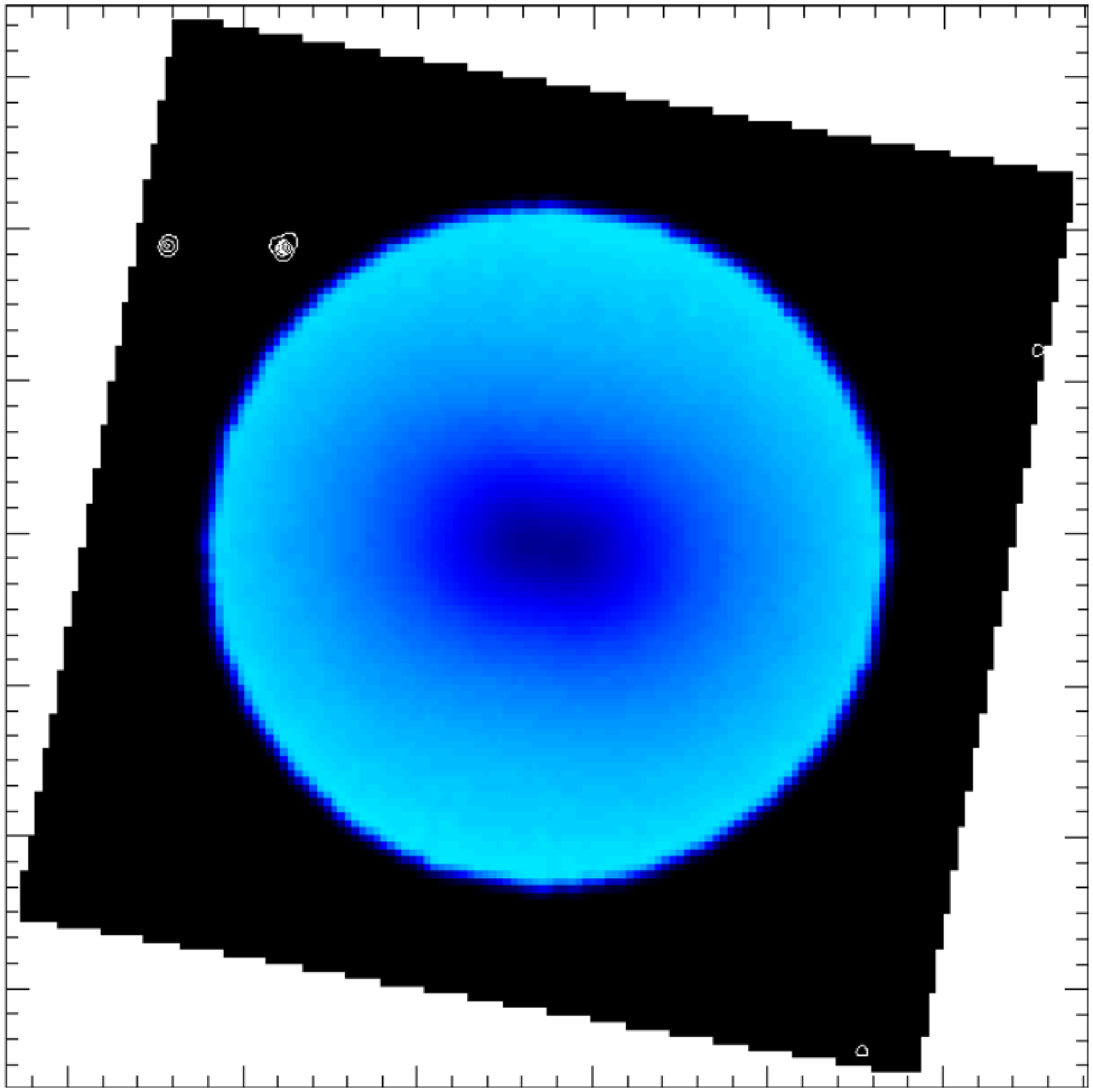}
\includegraphics[angle=0,width=55mm]{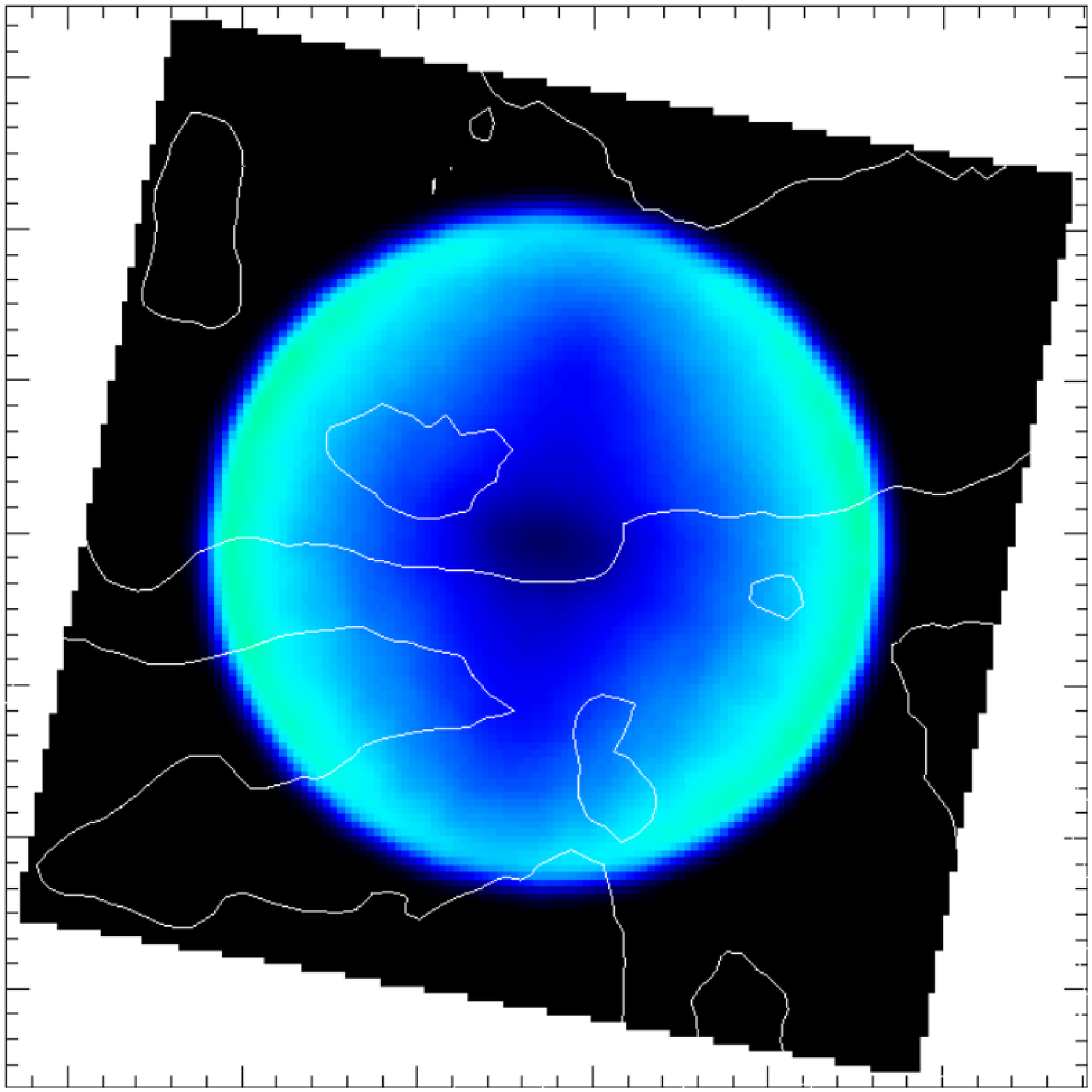}
\includegraphics[angle=0,width=55mm]{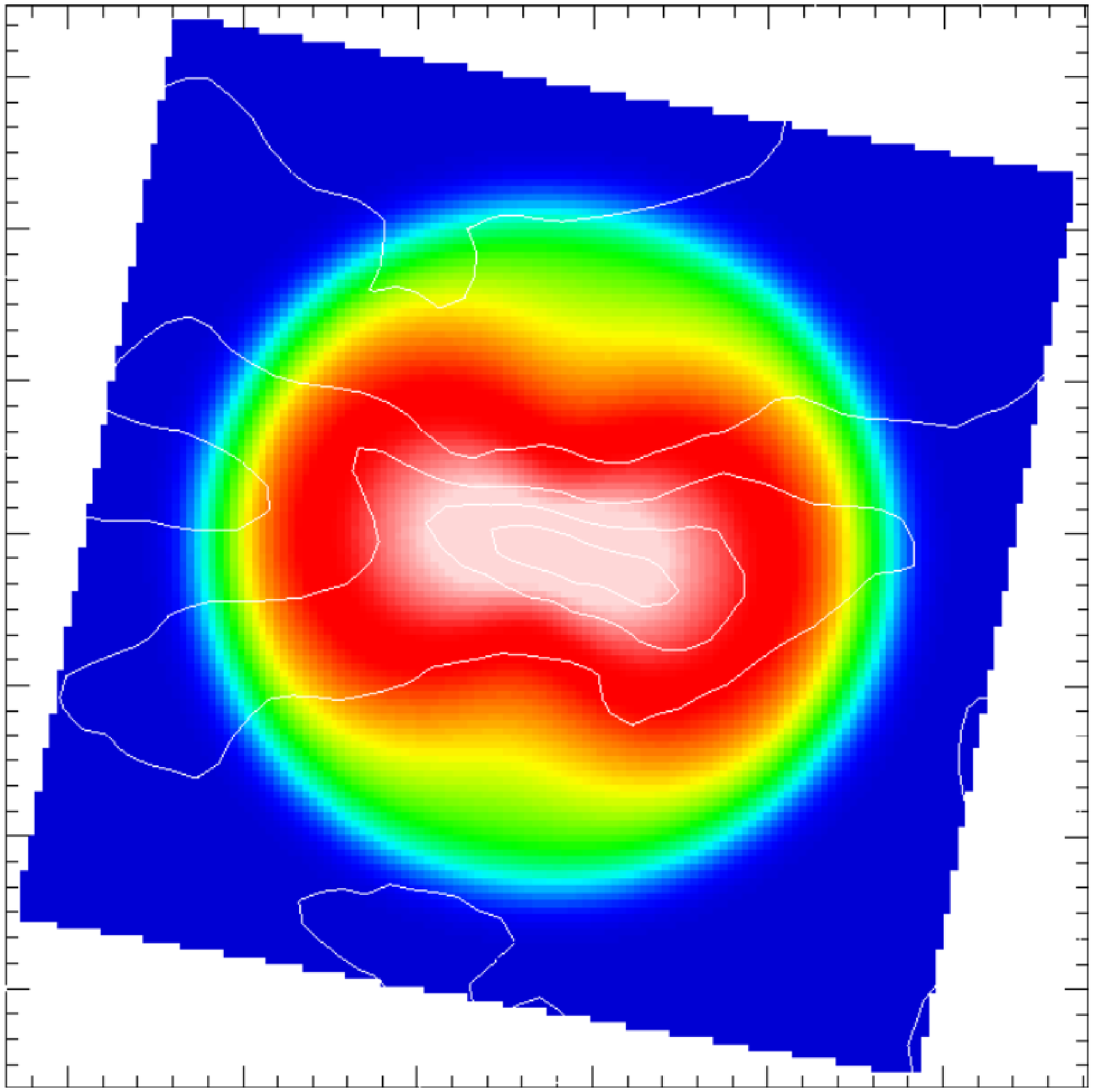}
\includegraphics[angle=-90,width=55mm]{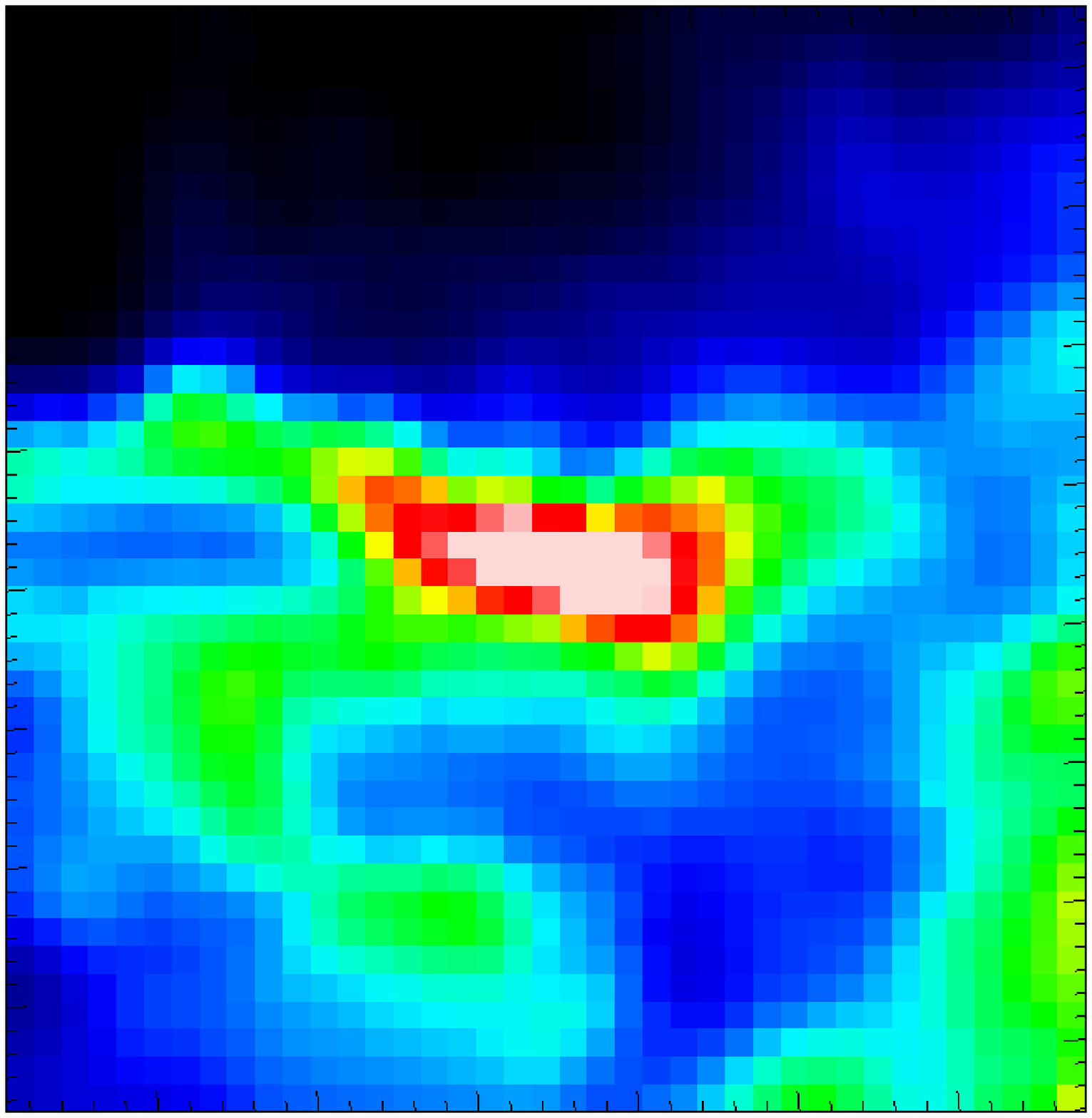}
\includegraphics[angle=-90,width=55mm]{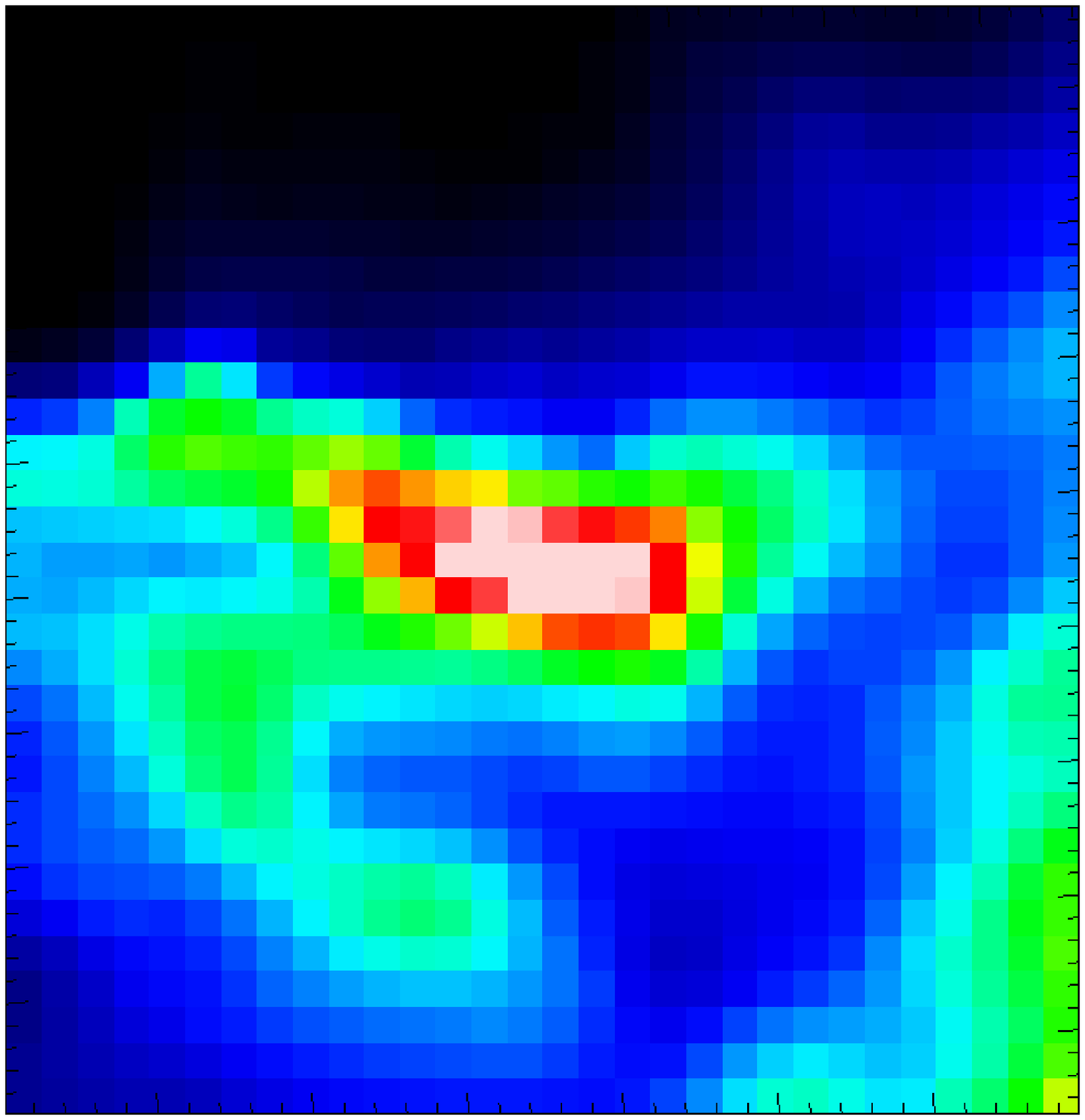}
\includegraphics[angle=-90,width=55mm]{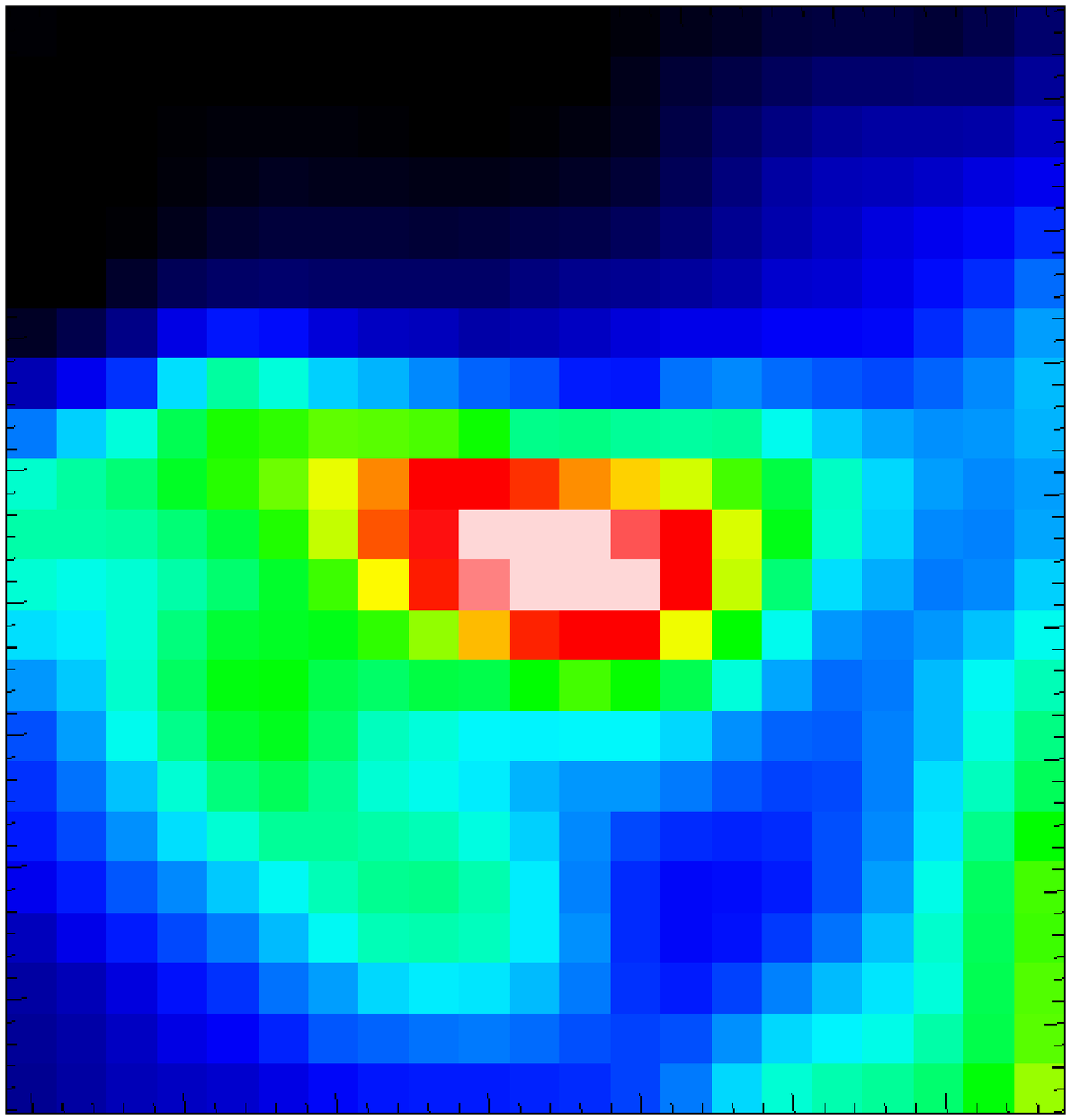}
\includegraphics[angle=0,width=55mm]{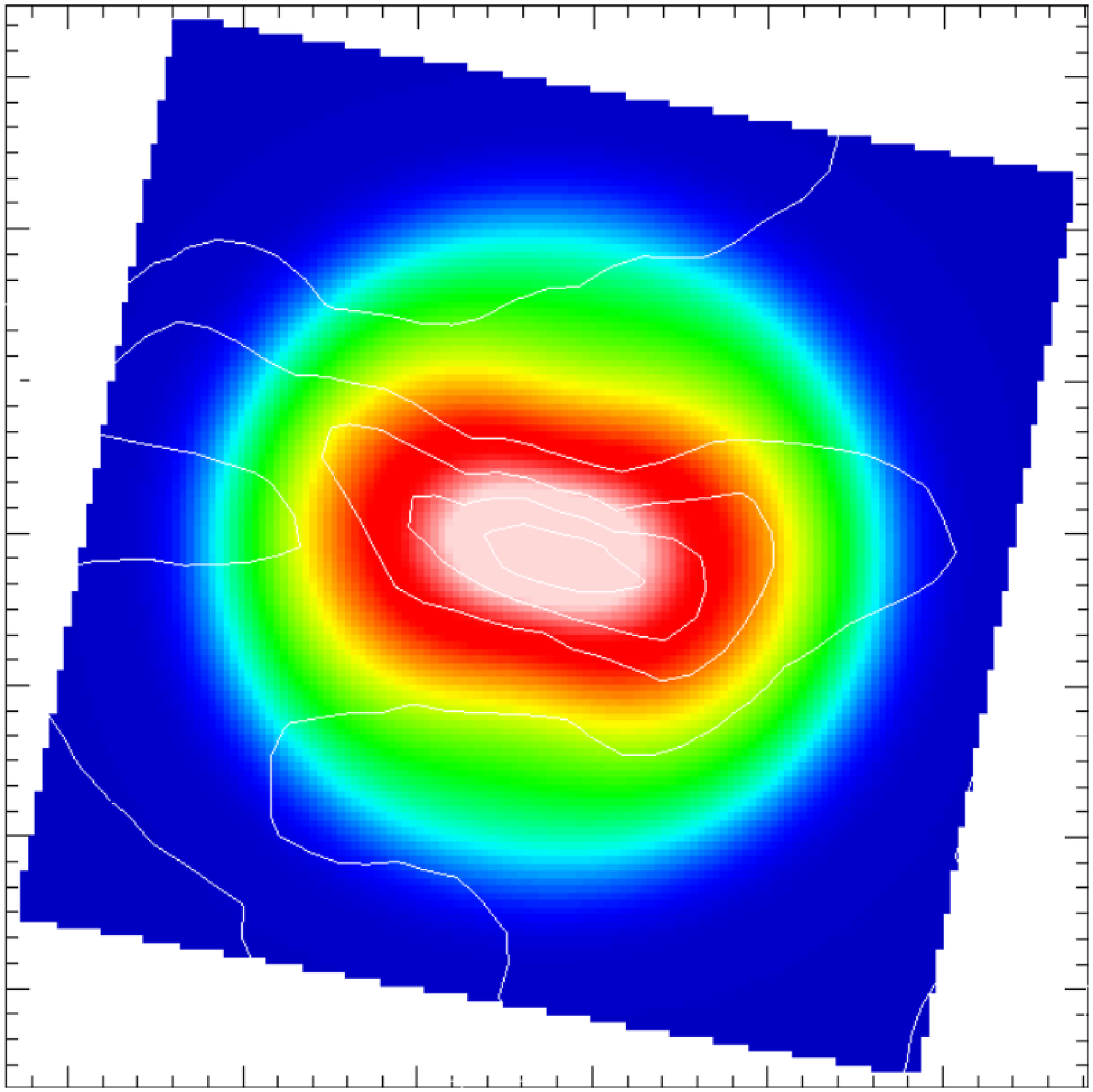}
\includegraphics[angle=0,width=55mm]{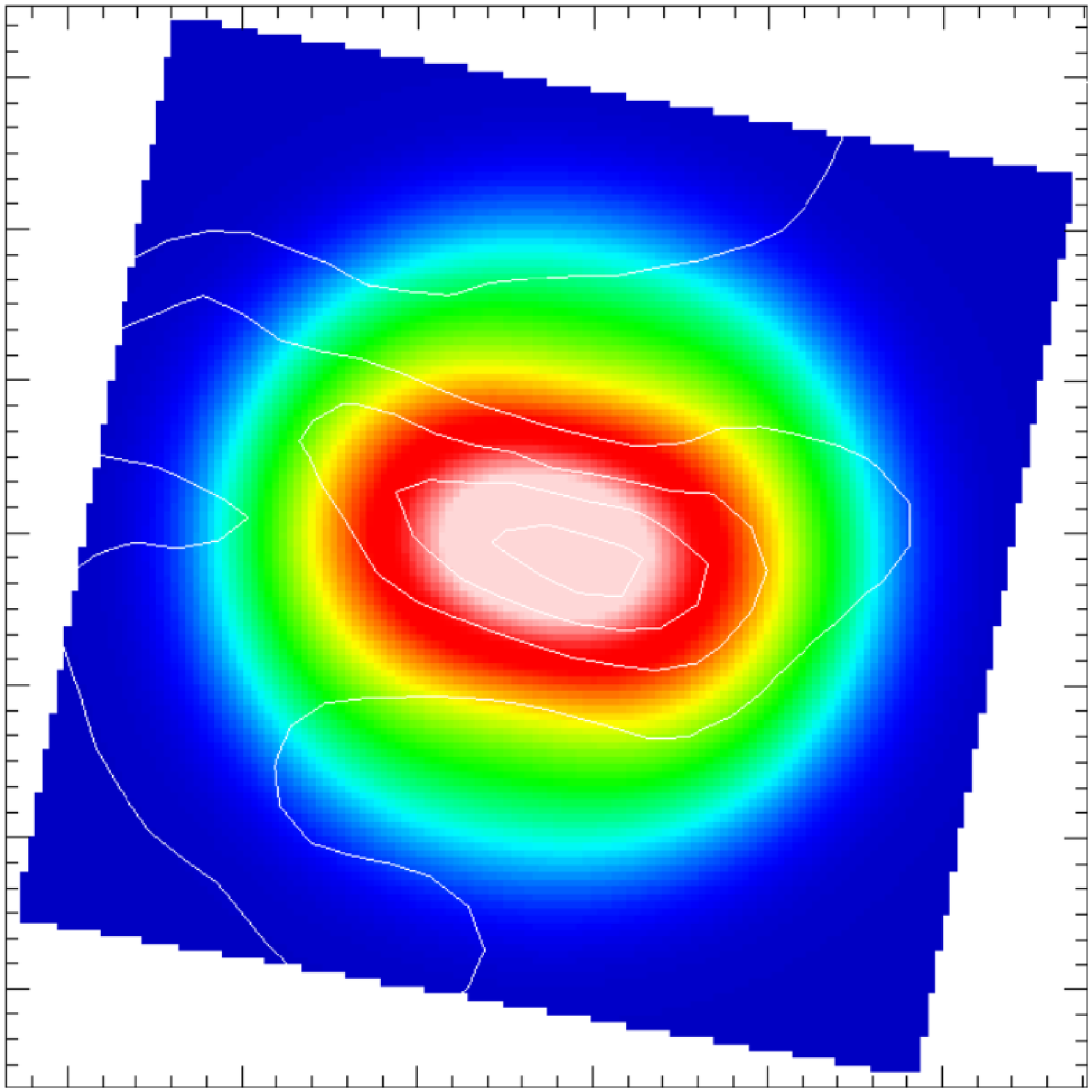}
\includegraphics[angle=0,width=55mm]{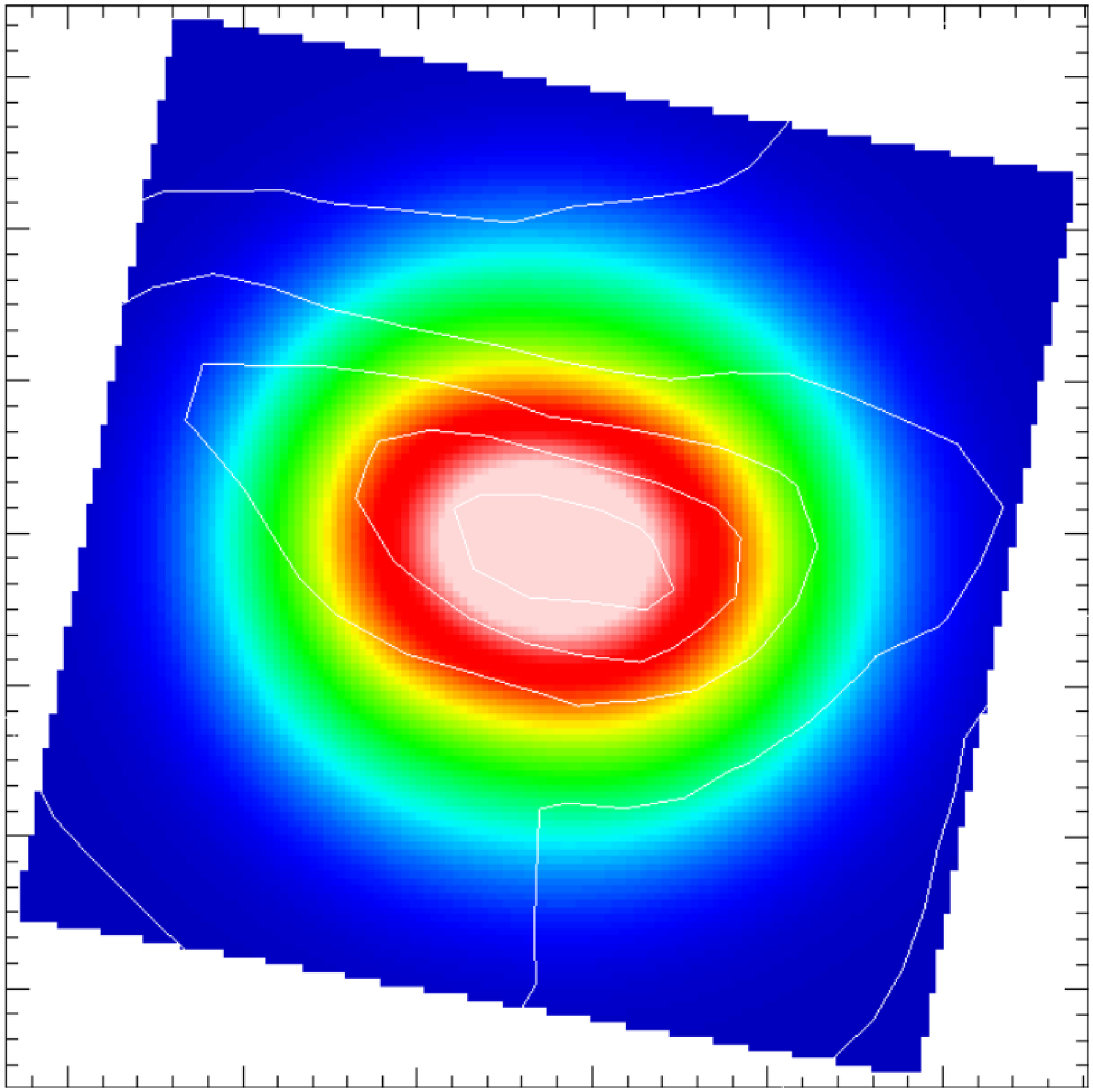}
\caption{305.798-0.097 - Upper row: Observational images of the infrared dark core at (left-right) \textit{Spitzer} 8\,$\mu$m, 
PACS 70\,$\mu$m and PACS 160\,$\mu$m. Second row: Modelled images at 8\,$\mu$m, 70\,$\mu$m and 160\,$\mu$m. Third row: 
Observational at SPIRE 250, 350 and 500\,$\mu$m. Lower row: Modelled images at 250, 
350 and 500\,$\mu$m. All images are $\sim$0.05$\times$0.05$^{\circ}$ in size, equivalent to 2.56$\times$2.56\,pc.} 
\label{irdc305798}
\end{figure*}

\end{document}